\documentstyle[prl,aps,floats,epsf]{revtex}
\font\msbm=msbm9  scaled \magstep 1    % Please, use this definition of
\def\kappa{\hbox{\msbm\char"7B}}       % {\kappa = \varkappa} in the paper
\begin{document}
\draft
\tighten
\twocolumn[\hsize\textwidth\columnwidth\hsize\csname
@twocolumnfalse\endcsname
\title{Self-consistent solutions of Ginzburg--Landau equations and \quad\quad\quad\quad\quad\quad
superconducting edge-suppressed states in magnetic field}
\author{G.F. Zharkov, V.G. Zharkov and A.Yu. Zvetkov}
\address{P.N. Lebedev Physics Institute, Russian Academy of Sciences,
 Moscow, 117924, Russia}
\date{August 10, 2000}
\maketitle
\begin{abstract}
Self-consistent solutions of the Ginzburg-Landau system of       %nonlinear
equations, which describe the behavior of the order parameter $\psi$, and the
magnetic field distribution $B$, in a long superconducting cylinder of finite
radius $R$, in external magnetic field $H$, when a vortex line, carrying $m$
flux quanta, is situated on the cylinder axis (a giant $m$-vortex state),
are studied by using numerical method. If the field $H$ exceeds some critical
value $H_s^{(m)}$, the giant $m$-vortex state solution becomes unstable (in
response to small fluctuations in its shape) and the solution passes to a
new stable edge-suppressed form. The quantum number $m$ in this state does
not change, but the order parameter diminishes by a jump (almost to zero) near
the cylinder surface; however, superconductivity remains in the deep, at some
distance from the cylinder axis. This edge-suppressed state exist in the
fields $H_s^{(m)}<H<H_c^{(m)}$, where $H_c^{(m)}$ is the field, in which the
second order phase transition into the normal state occures. (In the case of
large radii $R$ and finite $m$ the field $H_c^{(m)}$ coincides with the
critical field $H_{c2}$, in which the superconductivity vanishes in the bulk.)
If the magnetic quantum number $m$ is large, the
edge-suppressed state degenerates into the usual state of surface
superconductivity, and may survive up to the field $H_{c3}=1.69H_{c2}$.
The magnetic moment, the total magnetic flux, the Gibbs
free energy and other characteristics of the system are found, as functions of
the field $H$ and temperature $T$, for different radii $R$, quantum numbers $m$
and parameters  $\kappa$ of the Ginzburg--Landau theory. The intervals of $R$,
$T$, $\kappa$, $m$ are found, where the edge-suppressed solutions exist.
The giant $m$-vortex states exist in both type-I and type-II superconductors,
but the edge-suppressed solutions are possible only in type-II superconductors
(with $\kappa>1/\sqrt{2}$). The paramagnetic effect in mesoscopic samples and
also the possible connection of the theory and experiment are shortly discussed.
\end{abstract}
\pacs{ }
\vskip 2pc ] % end \twocolumn[...]
\narrowtext

\section{Introduction}

The behavior of finite size superconductors in magnetic field was studied on
the base of nonlinear system of Ginzburg--Landau equations [1] in numerous
theoretical papers (see, for instance, [2--15]). The results of [2--15] were
used, in particular, to explain some anomalies, observed in a number of
experiments with small size superconductors, placed in external magnetic field
[16--24].

In our recent paper [25], by using numerical methods, the self-consistent
solutions of nonlinear Ginzburg-Landau equations were studied, for a long
superconducting cylinder of finite radius in external axial magnetic field,
with no vortices inside the cylinder ($m=0$). It was found, that type-II
superconducting cylinder, which is in the Meissner (vortex-free) state, upon
increasing the field $H$, may pass by a jump into a special (also vortex-free)
edge-suppressed state. The order parameter $\psi$ in this state is strongly
suppressed (practically to zero) in the vicinity of the cylinder surface,
however, the superconductivity survives near the specimen center, where
$\psi$ remains finite. Such edge-suppressed solutions were not studied
previously [1--15].

Apart from the the Meissner-state ($m=0$) more general one-dimensional
solutions exist, when the vortex line, carrying arbitrary number of flux
quanta ($m=1,\,2,\,3,\dots$), is situated on the cylinder axis. Such solutions
(which are called "the giant-vortex states") have been extensively studied in
the literature ([2-30]) and used in the interpretation of experiments. The shape
of the giant-vortex changes smoothly with the external field $H$. As is shown
in the present paper, for a cylinder of finite radius, the giant-vortex state
(with vorticity $m>0$) becomes unstable, when the field $H$ exceeds some
critical value $H_s^{(m)}$. After that, the giant-vortex solution transforms
its shape by a jump and acquires the "edge-suppressed" form with the same
vorticity $m$. Such edge-suppressed states are characterized by a strongly
suppressed value of the order parameter $\psi$ near the cylinder boundary, and
are analogous to the edge-suppressed (or, rim-suppressed) states, found in [25]
for the case $m=0$. Previously, such edge-suppressed modifications of the
giant-vortex states were not studied in detail, though some evidence of the
edge-suppression was noticed also in [2,3,7].

We should stress, that the edge-suppressed state is noting else, but the usual
giant $m$-vortex solution in a field $H>H_s^{(m)}$. If the field $H$ is
increased further, the edge-suppressed solution continues to vary its form, the
maximal value of the order parameter $\psi_{max}^{(m)}$ gradually diminishes,
and for $\psi_{max}^{(m)}\ll 1$ the edge-suppressed solution may be expressed
in terms of the Kummer functions (see, for instance, [6,8]).

As is shown below, the edge-suppressed states can be formed only in type-II
superconductors, with $\kappa>1/\sqrt{2}$, where $\kappa$ is the
Ginzburg-Landau parameter. In type-I superconducting cylinder, with
$\kappa<1/\sqrt{2}$, the edge-suppressed states do not exist.

It is known, that for $m>0$ the order parameter vanishes in the vicinity of
the cylinder axis according to the law $\psi(r)\sim r^m$ [29]; if the field
increases, $\psi$ vanishes also near the surface (where the edge-suppressed
state forms), so, the superconductivity persists only at some distance from
the cylinder axis. If the magnetic quantum number $m$ is increased, the region,
where the order parameter $\psi(r)\ne 0$, shifts more and more toward the
cylinder surface. For a fixed cylinder radius ($R=$const, $\kappa>1/\sqrt{2}$)
there exists some maximal value $m=m_{max}$, for which the edge-suppressed state
degenerates into the usual state of the surface superconductivity [26--30].

The transition of a superconducting cylinder into the edge-suppressed state
may be accompanied by jumps in the dependencies versus $H$ of such quantities,
as: the specimen magnetic moment ($-4\pi M$); the total magnetic flux, captured
inside the system ($\Phi_1$); the system Gibbs free energy ($G$); and by
peculiarities in the behavior of other parameters, which characterize the
superconducting state. These topics are discussed below in detail. No immediate
comparison of the theory with the experiment is possible, because the model
case of an infinitely long cylinder approximates the real samples geometry
[16-24] very remotely. However, a number of qualitative predictions, which
follow from this model, may be used for interpreting some of the peculiarities,
observed in the experiments with mesoscopic samples. In particular, the
controversial paramagnetic Meissner effect in superconductors is shortly
discussed below, basing on the Ginzburg--Landau theory approach.

The paper is organized as follows. In Sec. 2, the basic equations and the
boundary conditions of the problem are written. The numeric algorithm, used
to find the self-consistent solutions of nonlinear system of equations, is
shortly described. Sec. 3 contains the results of numerical calculations.
We are forced to present our results by a large number of graphics, which
illustrate different details of the system behavior in this many parameter
problem (the solutions depend on $m$, $\kappa$, $R$, $H$, the specimen
temperature $T$, the critical temperature $T_c$ and the coherence length
$\xi_0$). The necessary comments accompany the presentation of the material.
Sec. 4 contains a short resume and discussion of the results.

\section{The setting of the problem}

Consider long superconducting cylinder of radius $R$ in the external magnetic
field $H$, which is parallel to the cylinder element. The basic system of the
Ginzburg-Landau equations [1] is of the form
$$ {\rm rot\, rot}{\bf A}=
{{4\pi} \over c}{\bf j}_s,\quad { {4\pi} \over c}{\bf j}_s = { {\psi^2} \over
{\lambda^2} } \left( { {\phi_0}\over {2\pi} } \nabla\Theta - {\bf A} \right),
$$
$$ \nabla^2 \psi - \left( \nabla\Theta - {2\pi \over \phi_0}{\bf A}
\right)^2 \psi + { 1 \over {\xi^2} }( \psi - \psi^3 )=0,
$$
where ${\bf A}$ is the vector-potential of the magnetic field
(${\bf B}={\rm rot} {\bf A}$), ${\bf j}_s$ is the current density inside the
superconductor, $\lambda$ is the field penetration depth, $\xi$ is the
coherence length, $\lambda=\kappa\xi$. The order parameter, in a general
case, is written as $\Psi=\psi e^{i\Theta}$, where $\psi$ is the modulus and
$\Theta$ is the phase of the order parameter. From the single-valuedness
of $\Psi$ the condition follows
$$ \oint_C
\nabla\Theta d{\bf l}=2\pi m ,
$$
where the contour $C$ embraces the vortex axis, $m$ is integer (the
topological invariant, or fluxoid, or vorticity), which shows, how many
vortices may be present inside the contour $R$.

In the cylidrical system of co-ordinates $r, \varphi, z$, with $z$ axis directed
along the cylinder element [when the vector-potential has only one component,
${\bf A}={\bf e}_\varphi A(r)$], these equations may be written in the
dimensionless form
$$
{ {d^2U} \over {d\rho^2} } - {1\over \rho}{ {dU}\over {d\rho} } -
\psi^2 U=0,                                                    \eqno(1)
$$
$$ { {d^2 \psi} \over {d\rho^2} } + {1\over\rho}
{{d\psi}\over{d\rho}} + \kappa^2 (\psi - \psi^3) - { {U^2 } \over {\rho^2}
}\psi =0.                                                     \eqno(2)
$$
Here, instead of the dimensioned potential $A$, field $B$ and current $j_s$,
the dimensionless quantities $U(\rho ),\, b(\rho )$ and $j(\rho )$ are
introduced:
$$ A={ {\phi_0} \over {2\pi\lambda} }{ U + m \over \rho },\quad
B={ {\phi_0} \over {2\pi\lambda^2} }b,\quad
b={1\over \rho}{{dU}\over{d\rho}},                              \eqno(3)
$$
$$
j(\rho)=j_s\Big/ { {c\phi_0} \over {8\pi^2\lambda^3} }=
-\psi^2 {U\over \rho},\quad  \rho = {r\over \lambda}.
$$
(The field $B$ in (3) is normalized by $H_\lambda=\phi_0/(2\pi\lambda^2)$,
with $b=B/H_\lambda$; instead of $H_\lambda$ one can normalize by
$H_\xi=\phi_0/(2\pi\xi^2)\equiv H_{c2}$ [28--30], or by thermodinamical
critical field $H_c=\kappa H_\lambda/\sqrt{2}$. The coefficents in (1)--(3)
would change accordingly.)

The magnetic flux, confined inside the contour of the radius $r$, is
$$
\Phi= \int {\bf B}d{\bf s}= \oint_{r}{\bf A}d{\bf l} =\phi_0 (U+m),\
U=U(\rho),\ \rho={r \over \lambda }.
$$
Thus, the potential $U(\rho)$, in the normalization adopted above, is
related to the flux $\Phi(\rho)$ by a simple formula
$\phi\equiv\Phi/\phi_0=U(\rho)+m$.

Because the magnetic flux through the vanishing contour is zero, and the field
$B|_{r=R}=H$, the following boundary conditions correspond to Eq. (1):
$$ U\big|_{\rho =0} = -m,\quad
 \left. { {dU}\over{d\rho} }\right|_{\rho =\rho_1}=h_\lambda,      \eqno(4)
$$
where $\rho_1=R/\lambda$, $h_\lambda=H/H_\lambda$,
$H_\lambda=\phi_0/(2\pi\lambda^2)$.

As to the Eq. (2), we shall take the usual boundary condition on the external
surface [1]: $d\psi / d\rho |_{\rho =\rho_1}=0$. The order parameter at the
center is either maximal (if $m=0$), or zero (if $m>0$, see, for instance,
[29]), thus, the following boundary conditions correspond to Eq. (2):
$$
\left. {d\psi \over d\rho} \right|_{\rho =0} =0, \quad
\left. {d\psi \over d\rho} \right|_{\rho=\rho_1} =0 \quad (m=0), \eqno(5)
$$
$$
\psi|_{\rho=0}=0,\quad
\left. { d\psi \over d\rho} \right|_{\rho=\rho_1} =0 \quad (m>0). \eqno(6)
$$

The magnetic moment (or, magnetization) of the cylinder, related to the unity
volume, is
$$
{M\over V}={1\over V}\int  { B-H \over 4\pi }dv =
{ B_{av}-H \over 4\pi },                                         \eqno(7)
$$
$$
B_{av}={1\over V}\int B({\bf r})dv={1\over S}\Phi_1,
$$
where $B_{av}$ is the mean field value inside the superconductor,
$\Phi_1=\Phi(R)$, $S=\pi R^2$. In normalization (3), denoting
$\overline{b}=B_{av}/H_\lambda$, $h_\lambda=H/H_\lambda$,
$M_\lambda=M/H_\lambda$, one finds from (7):
\begin{eqnarray*}
4\pi M_\lambda=\overline{b}-h_\lambda, \ \
\overline{b}={2\over\rho_1^2}(U_1+m), \ \
\phi_1={\Phi_1 \over \phi_0}=U_1+m,\\
U_1=U(\rho_1),\quad \rho_1={R\over\lambda}.\qquad\qquad\qquad  (8)
\end{eqnarray*}

For the difference of Gibbs free energies of the system in superconducting
and normal states, $\Delta G=G_s-G_n$, it is convenient to use the exact
expression:
$$
\Delta G={\cal F}_{s0}-{1\over 2}MH+{\phi_0 Z \over 8\pi} m(B(0)-H), \eqno(9)
$$
$$
{\cal F}_{s0}={ H_c^2 \over 8\pi }\int \left[
\psi^4\!-2\psi^2\!+\xi^2\!\left( {d\psi \over dr} \right)^2 \right] dv,
H_c={\phi_0 \over 2\sqrt{2}\pi\lambda\xi},
$$
where ${\cal F}_{s0}$ corresponds to the superconductor condensation energy,
$M$ is the cylinder magnetic moment, $Z$ is the cylinder length, $m$ is the
magnetic quantum number, $B(0)$ is the magnetic field at the cylinder axis,
$H$ is the external magnetic field. The expression (9) follows from the
general formula, obtained in [31] for the free energy of a hollow
superconducting cylinder, if the radius of a hollow is put to zero (see also
[32,33]). Using normalization (3), and also (8), one finds from (9) the
normalized expression
$$
\Delta g=\Delta G\Big/ \left( { H_{c{\rm m}}^2 \over 8\pi } V \right)=
g_0-{8\pi M_\lambda \over \kappa^2} h_\lambda +{4m \over \kappa^2}
{b(0)-h_\lambda \over \rho_1^2},                          \eqno(10)
$$
$$
g_0={2\over\rho_1^2} \int_0^{\rho_1}
\rho d\rho \left[ \psi^4-2\psi^2+{1\over\kappa^2} \left(
{d\psi \over d\rho} \right)^2 \right].
$$
The expressions (3),(8),(10) will be used in Sec. 3 for calculating the
corresponding quantities.

We remind, that the field penetration depth $\lambda$ and the coherence length
$\xi=\lambda/\kappa$ depend on temperature. Thus, the expressions above depend
implicitly on temperature, and, formally, are valid for arbitrary values of
$T$. (Though, the Ginzburg-Landau equations themselves are applicable only in
the limit $T\to T_c$ [34,28--30], when the expression
$\xi(T)=\xi_0/\sqrt{1-T/T_c}$ may be used for the coherence length).

It is appropriate to make here a comment on the iteration procedure, we used
to obtain the self-consistent solution of the system of equations (1)--(6).
First, some trial function $\psi(\rho)$ was chosen, and the solution for
$U(\rho)$ was found from Eqs. (1), (4). Then, this function $U(\rho)$
was introduced into Eq. (2), which was solved with account of the boundary
conditions (5), (6) and new function $\psi(\rho)$ was found. Further, Eq.(1)
was solved again, and all the procedure was repeted, until the functions
$\psi(\rho)$ and $U(\rho)$ ceased to change, representing the self-consistent
solution of the system (see also [25,35]). However, the numeric algorithm, used
in our calculations, does not allow to find the unstable "superheated" and
"supercooled" states in type-I superconductors (with $\kappa<1/\sqrt{2}$),
where other methods are required (see [32,36,37]).

The question of stability requires an additional illucidation.
Alongside with the centrally symmetric giant-vortex states, the so-called
multi-vortex solutions of the same vorticity $m$ exist. The multi-vortex states
correspond to the case, when a single giant vortex (with vorticity $m$) decays
into a number of vortices (with vorticities $m_i,\, \sum m_i=m$). These vortices
can be arbitrary positioned on the cylinder cross-section, so the multi-vortex
state may be no longer centrally symmetric. The question, which of these
multi-vortex states possesses the minimal free energy, is important, because
the physical system is usually assumed to occupy the state with the smallest
energy (the equilibrium ground state), and all other states are considered as
energetically unstable (or metastable). The free energy calculations to find
the equilibrium multi-vortex configuration is a difficult problem, which
requires the specific methods of investigation. (See, for instance, the theory
[30], where the triangular lattice of single vortices is found as an
equilibrium state for the case of infinite superconductor, placed in the field
$H\sim H_{c1}$. The numeric calculations for mesoscopic suprconducting discs
may be found in [6--15], where a simplified model of the vortex distribution
is considered. To find the equilibrium vortex configuration in a general case
of a finite-size superconductor, placed in arbitrary magnetic field $H$, is a
problem as yet unsolved.)

In the present work a limited task was set: to describe the formal properties
of the giant-vortex states, which were not reported previously (in particular,
the edge-suppressed transitions and the jumps in the magnetization curves with
a fixed vorticity $m$). The general problem of energetic stability and of the
equilibrium vortex configuration is left outside the scope of the present
investigation. Some of the edge-suppressed states, which are described below,
might in reality be energetically metastable and decay into an equilibrium
ground state configuration. However, as is well known, the physical system not
necessarily must be in equilibrium, but (with some probability) may also occupy
the excited metastable states. The giant-vortex and the edge-suppressed
solutions provide the system this opportunity, so the behavior of a
superconductor in such states might be of some interest.

The results of our calculations are presented below.

\section{The numerical results}

The results of the investigation are illustrated below by a number of
graphics, which detalize various aspects of this multy-parameter task.
(Some of these results can be obtained by the methods, used in the papers,
cited above. However, for the uniformity of presentation, we prefer to
reproduce them by our method.)

Evidently, the superconductivity would be destroyed by a large external field
$H$ with subsequent transition to the normal state. In Fig. 1($a$) the phase
diagram is depicted (for $\kappa=2$ and different $m$), which separates
the normal ($n$) and superconducting ($s$) states on the plane of variables
$R_\lambda=R/\lambda$ and $h_\xi=H/H_\xi$. The phase diagram is a many-sheet
surface, each sheet corresponds to the fixed value of the vorticity $m$.
The sheet $m=0$ is the lowest laying (according to the increasing $m$),
the sheet $m=1$ lays above it, still more above lays the sheet $m=2$, and so
on. (It is sufficient to consider only the values $m\ge 0$, because the
results for $m<0$ are easily reproducible with account of the evident symmetry
$(m,H)\iff (-m,-H)$.)

On each of $m$-sheets of the phase surface there is a phase curve ($h_c$,
which may be marked by the index $m$), which separates the region of values
$(R_\lambda,h_\xi$), where the superconducting solutions exist (i.e.
$\psi>0$), from the region of normal state solutions ($\psi\equiv 0$). For
a bulk superconductor ($R_\lambda\gg 1$) the superconducting state boundary
(in the case of type-II superconductor) lays at $h_\xi=H/H_\xi=1$, i.e. at
$H=H_{c2}=\phi_0/(2\pi\xi^2)$ [28--30] ($H_{c2}\equiv H_\xi$). As can be seen
from Fig. 1($a$), for small $R_\lambda$ the superconducting states (with
different values of $m$) are also possible in the fields, exceeding the
critical value $H_{c2}$ (i.e. for $h_\xi>1$). For fixed $R_\lambda$ and $m$
there exists a maximal field $h_\xi=h_c^{(m)}$, above which the superconducting
solution is impossible. As is well known, the maximal field, at which the
superconductivity is still possible (for a bulk specimen with $R_\lambda\gg 1$)
is $h_\xi=1.69$ (i.e. $H=H_{c3}$, where $H_{c3}=1.69H_{c2}$ is the field, at
which the surface superconduvtivity nucleates in macroscopic specimens
[26--30]). As is evident from Fig. 1($a$), for small $R_\lambda$, the
superconductivity is also possible in the fields, exceeding $h_\xi=1.69$.
[The analogous phase diagram in the case of type-I superconductor (with
$\kappa=0.5$, $m=0,\,1,\,2$) is depicted in Fig. 1($b$).]

Inside each of the superconducting regions in Fig. 1($a$) there exists a
singular line $h_s$, which is depicted, for clarity, on a separate phase
$m$-sheet (see Fig. 2). In Fig. 2($a$) the phase sheet $m=0$ (for $\kappa=2$)
is depicted. The solid line $h_c$ corresponds to the superconducting state
boundary. The letter $s$ marks the region, where the superconducting state
exists ($\psi\ne 0$), the letter $n$ denotes the normal state ($\psi=0$).

To every point($R_\lambda,h_\xi$), which lays inside the superconducting region
in Fig. 2($a$), corresponds some solution of Eqs. (1)--(6). When this point is
shifted, the solution changes accordingly. The central part of the drawing
(which lays in the withinity of $h_\xi\sim 0$) corresponds to the
superconducting Meissner state. The order parameter here is $\psi\sim 1$, the
external field $H$ is almost completely screened out and penetrates only to the
distance $\sim\lambda$ from the superconductor surface. The shape of the
Meissner solution depends on $H$ only slightly.

There exists a critical line $h_s$ in Fig. 2($a$), near which the Meissner
solution ($\psi\sim 1$) starts to change strongly. The character of this
transformation depends on the cylinder radius $R_\lambda$. If the representation
point $(R_\lambda,h_\xi)$ crosses the the critical line $h_s$ above the point
$R_\Delta$ ($R_\lambda>R_\Delta$), the static Meissner solution becomes
absolutely unstable relative small variations in its form, and transforms by a
jump into new stable (static) superconducting solution with the same $m$. In
this new state (which may be called as "edge-suppressed", or "rim-suppressed"
[25]) the order parameter $\psi(r)$ is strongly suppressed in some layer, which
is situated near the superconductor surface, while the magnetic field penetrates
this layer practically without screening (see Fig. 2 in [25] for $m=0$ and also
Figs. 3--5 below).

The jump transformation of the order parameter shape is acompanied by jumps in
the magnetization and in other physical characteristics (see Fig. 3 below).
The amplitude of these jumps diminishes with $R_\lambda$ diminishing, and for
$R_\lambda=R_\Delta$ the jump amplitude (see the broken vertical lines
in Fig. 3) vanishes. Thus, for $R_\lambda>R_\Delta$ the transition of the
Meissner state ($m=0$) to the edge-suppressed form ($m=0$) is the first order
phase transition (by a jump).

For $R_\lambda<R_\Delta$ the Meissner solution also changes its form
significantly, when the line $h_s$ is crossed, but this transformation happens
smoothly, without jump (a second order phase transition to the edge-suppressed
form). In the case $R_\lambda=R_\Delta$ the jump amplitude is zero, so the
first and second order phase transitions to the edge-suppressed state become
indistinguishable. (The meaning of the points $R_w,\,R_0,\,R_\Delta$ in Figs.
2($a,d$) will be explained later.)

As was noted, to every point of $s$-region in Fig.2($a$) corresponds some
solution of Eqs. (1)--(6) for functions $\psi$ and $U$, which describe the
superconducting state. When the representation point ($R_\lambda,h_\xi$)
moves, this state changes. Let us trace in more details, what happens, if the
parameter $R_\lambda=R/\lambda$ remains constant (i.e. the temperature remains
constant), but the parameter $h_\xi=H/H_\xi$ changes (i.e. only the field $H$
changes).

In Fig. 3 are depicted as functions of $H$:
($a$) -- the system magnetic moment ($-4\pi M_\lambda$) (8);
($b$) -- the average magnetic field $\overline{b}$ [or, the total magnetic
flux inside the cylinder, $\phi_1={1\over 2}\rho_1^2 \overline{b}$ (8)];
($c$) -- the difference of the system free energies, $\Delta g$, (10);
($d$) -- the magnetic field magnitude at the cylinder axis, $b_0$;
($e$) -- the maximal value of the superconducting order parameter,
$\psi_{max}$;
($f$) -- the order parameter value at the cylinder surface, $\psi_1$.
Fig. 3 corresponds to the case $m=0$, $\kappa=2$, $R_\lambda=5$.
[The superconducting region in Fig. 2($a$) is crossed along the line
$R_\lambda=5$.] The values of $h_\lambda=\kappa^2 h_\xi$ are plotted on the
horizontal axis. By open circles the points are marked, which deserve a
special commentary.

As is seen from Fig. 3($a$), the magnetic moment ($-4\pi M_\lambda$) increases
linearly for small $h_\lambda>0$ ($h_\lambda=0$ at the point {\it 4}). The
average magnetic field inside the cylinder, $\overline{b}$, is essentially
reduced, in comparison to the normal state (see the dotted line $n$ in Fig.
3($b$)). (Analogously behaves the total magnetic flux, confined inside the
cylinder, $\phi_1$). The initial linear part reflects the Meissner effect,
it corresponds to the external field expulsion from the superconductor interior.
If the field $H$ is increased further, the magnetic moment increases up to the
point {\it 5}, where the magnetization diminishes by jump, and takes the
value {\it 6}.

The analogous jumps are present in other quantities, depicted in Figs.
3($b-f$).  The value of the order parameter at the surface, $\psi_1$ (Fig.
3($f$)), changes especially strongly: from $\psi_1=0.706$ at the point {\it 5}
(where $h_\lambda=1.6837$), to $\psi_1=0.087$ at the point {\it 6} (where
$h_\lambda=1.6838$). The jump in the free energy $\Delta g$ (Fig. 3($c$)) is
not seen in this scale.

When the external field $h_\xi$ is increased beyond the point {\it 6}, the
maximal value of the order parameter, $\psi_{max}$, starts diminish quickly.
This is acompanied by a fast drop in the "tail" of the magnetization. (This
"tail" lays between the points {\it 6} and {\it 7} in Fig. 3($a$) and
represents the edge-suppressed solution, $e_0$.) When the field $h_\xi$ reaches
the point {\it 7}, the order parameter vanishes completely, by the second order
phase transition to the normal state ($\psi_{max}\to 0$).

Fig. 4 ($m=0$, $\kappa=2$) demonstrates, what is going on, when the cylinder
radius $R$ is diminishing. [The superconducting region in Fig. 2($a$) is
crossed along the line $R_\lambda=3$.] Now, there are no jumps on the
magnetization curve (Fig. 4($a$)), which were represented by dashed lines in
Fig. 3($a$). But, the points of inflection {\it 2} and {\it 4} remain, where
the magnetization curve has the maximal derivative. The corresponding points
in Figs. 4($b,c$) are marked by the same numerals {\it 2} and {\it 4} as in
Fig. 4($a$). Evidently, the solution transforms smoothly, when the field
$h_\lambda$ passes the critical value $h_s$ (the point {\it 4}). This point
corresponds approximately to the value $R_\Delta$ in Fig. 2($a$), where the
derivative of the magnetzation curve is infinite.

In Fig. 5 the case $m=0$, $\kappa=2$, $R_\lambda=1$ is presented.
[The superconducting region in Fig. 2($a$) is crossed along the line
$R_\lambda=1$.] Here, the inflection point {\it 4} ($h_\lambda=5.91$) on the
curve ($-4\pi M_\lambda$) practically coincides with the termination point
{\it 5} ($h_\lambda=5.96$). This situation corresponds to the point $R_w$ in
Fig. 2($a$) (where $h_s$ and $h_c$ merge), and where the tail of the
magnetization disappears (i.e. the width of the tail vanish). For
$R_\lambda<R_w$ the second order phase transition to the normal state occures,
and the magnetization curve has no tail.

Evidently, the picture, presented by Fig. 2($a$) ($m=0$), is symmetric relative
the axis $h_\xi=0$. The analogous picture, which corresponds to the phase sheet
$m=1$ [Fig. 2($d$)], is asymmetric relative the axis $h_\xi=0$. The
edge-suppressed state exists only in the region $h_\xi>0$, and they do not
exist in the region $h_\xi<0$. This may be interpreted in the physical terms.

The phase sheet $m=1$ describes situation, when at the cylinder axis there
is a single vortex (its own field is assumed to have a positive sign), with the
order parameter vanishing on the axis: $\psi_0=0$. If the external field of the
same direction ($h_\xi >0$) is applyed, the magnetic field is additionally
pumped into the specimen. When the field reaches the critical value $h_s$, the
superconductivity is partly destroyed, and a new superconducting state forms,
with strongly suppressed order parameter $\psi_1$ at the cylinder surface (see
Fig. 6($f$) below). This state preserves up to the field $h_\xi=h_c^{(+)}$
(the right branch of the critical line, $h_c>0$), until the order parameter
vanishes finally everywhere. Thus, in the vicinity of the line $h_c^{(+)}$ in
Fig. 2($d$), the mechanism of the order parameter suppression by the external
field acts.

If the external field has the opposite sign [$h_\xi<0$, Fig. 2($d$)], the field
of the vortex ($m=1$) is gradual pumped out from the vortex interior. When the
external field reaches the critical value $h_\xi=h_c^{(-)}$ (the left branch
of the critical line, $h_c<0$), the vortex magnetic field $b_0$ is completely
sucked out from the superconductor (see Fig. 6($d$)). When this happens, the
vortex field at the axis vanishes (this means the jump transition from the
state $m=1$ to the state $m=0$); the order parameter also vanishes by a jump
(see Figs. 6($e,f$)). Thus, in the vicinity of the line $h_c^{(-)}$, the
mechanism of pumping out the field from the vortex interior acts. This explains
the asymmetric behavior of the curves $h_c^{(-)}$ and $h_c^{(+)}$ in
Fig. 2($d$).

Note also, that the curves $h_c^{(-)}$ and $h_c^{(+)}$ come up to the axis
$h_\xi=0$, having different derivatives (the fracture point at $h_\xi=0$ is
marked by an open circle in Fig. 2($d$); the corresponding value at
$R_\lambda$-axis is $R_0$).

Mention one more peculiarity of the curves, depicted in Fig. 2($d$). In the
case of small radii (for $R_\lambda<R_0$ and $m=1$) there exists a field
interval, where the superconducting state, which was impossible at smaller
$h_\xi$, becomes possible again for larger $h_\xi$. Here one has an
example of the so-called "reentry" superconductivity. In a cylinder of very
small radius ($R_\lambda<R_0$) the magnetic field of the vortex can not be
confined inside the superconductor and dissipates outside through the cylinder
surface. However, the imposition of the external field prevents the
dissipation of the vortex field and stabilizes the superconducting state. If
the external field is increased further, the superconductivity will be finally
destroyed. Thus, on the lowest part of the curve $h_c$ in Fig. 2($d$) there
exists a minimum at $R_\lambda=R_{\lambda\,min}$ (marked by an open circle).

If the parameter $\kappa$ is increased, the curves $h_s$ and $h_c$ behave
analogously to what is shown in Figs. 2($a,d$). But, if $\kappa$ diminishes
(see Figs. 2($b,e$)), the point, where the curves $h_s$ and $h_c$ merge,
raises up to larger $R_\lambda$, and the width of the region, where the
edge-suppressed state exists, diminishes. At $\kappa=1/\sqrt{2}$ the width of
this region vanishes, and the curves $h_s$ and $h_c$ merge into one indivisible
curve $h_c$. For $\kappa<1/\sqrt{2}$ (i.e. for type-I superconductors) the
edge-suppressed states do not exist. The behavior of the critical curves $h_c$
in the case of type-I superconductor is depicted in Fig. 2($c,f$) for different
$\kappa<1/\sqrt{2}$ ($m=0$ and $1$). One can see, that the superconducting
state with a vortex ($m=1$) at the cylinder axis, in principle, can exist in
both type-I and type-II superconductors.

The solution behavior in a case $m=1$, $\kappa=2$ [when the superconducting
region $s$ in Fig. 2($d$) is crossed along the line $R_\lambda=5$] is
illustrated in Fig. 6. There are also jumps, as in Fig. 3, but the curves are
asymmetric relative the axis $h_\xi=0$. In particular, on the magnetic moment
curve (Fig. 6($a$)), in a case $h_\lambda<0$ there is no "tail", which exists
in Fig. 3($a$) for $h_\lambda<0$ (between points {\it 1} and {\it 2}). This
asymmetry (as was already mentioned) may be explained by different mechanisms,
which act in destruction of the superconducting state $m=1$. If $H>0$, the
additional pumping of the external field into the superconductor occures, with
the subsequent suppresion of the order parameter by the field, followed by the
transition to the normal state (or, to the state $m\to m+1$). If $H<0$, the
field is pumped out from the vortex and the superconductor passes into the
vortex-free state $m=0$ (or, even into the normal state).

The analogous picture (for $m=1$, $\kappa=2$, $R_\lambda=3)$ is presented in
Fig. 7. In difference to Fig. 4 ($m=0$), the curves in Fig. 7 ($m=1$) have no
smooth "tails" for $h_\xi<0$ (to the left of points {\it 2}), but they
terminate by a jump at the points {\it 2}, where the field at the vortex axis,
$b_0$, vanishes, and the order parameter suffers a jump.

Fig. 8 corresponds to the case $m=1$, $\kappa=2$, $R_\lambda=1$. Notice, that
in Fig. 8($a$), at positive values $h_\lambda>0$ (in the interval between the
points {\it 3} and {\it 4}) the magnetic moment is positive
($4\pi M_\lambda>0$), which corresponds to the paramagnetic susceptibility. In
the interval between the points {\it 5} and {\it 6} the magnetic moment is
negative ($4\pi M_\lambda<0$), which corresponds to the diamagnetic
susceptibility. As can be seen from Fig. 8($b$), in the interval between
points {\it 3} and {\it 4}, the external field enhances the superconductivity
(because $\psi_{max}$ increases). In the interval between points {\it 4} and
{\it 5}, the external field suppresses the superconductivity (because
$\psi_{max}$ decreases).

The analogous dependencies are presented in Fig. 9 for type-I superconductor
($R_\lambda=5$, $\kappa=0.5$, $m=0,\,1$). The singular points are marked by an
open circles. There are no "tails" on the magnetization curves (Fig. 9($a$))
in both regions $h_\xi<0$ and $h_\xi>0$. This is characteristic for type-I
superconductors with $\kappa<1/\sqrt{2}$, when the destruction of the
superconductivity by the magnetic field proceeds in a jump, as the first
order phase transition. [This is valid, if $R_\lambda>R_\Delta$ in Fig. 1($b$).
If $R_\lambda<R_\Delta$, the phase transition to the normal state is of second
order, even in type-I superconductors; see Fig. 14 below.]

There are other features, common to all type-I superconductors. Thus, if the
family of curves, depicted in Fig. 2($e$) (for $m=0$ and different
$\kappa<1/\sqrt{2}$), is re-drawn on the ($R_\xi,h_\xi$)-plain, one obtains
the unified phase curve, Fig. 10($a$) ($m=0$). Notice also, that in the case
$m=0$ the curves with $\kappa<1/\sqrt{2}$ and $\kappa\ge 1/\sqrt{2}$ almost
coincide, so the phase curves in Fig. 10($a$) look like one universal function.
(For $h_\xi>1$ this function is $R_\xi\approx 2.8/h_\xi$).

Analogously, if the family of curves in Fig. 2($f$) is re-drawn on the
($R_\xi,h_\xi$)-plain, the Fig. 10($b$) emerges. One can see from this
figure, that for all $\kappa\le 1/\sqrt{2}$ there exists a minimal radius
$R_{\xi\, min}\approx 1.3$. For $R_\xi < R_{\xi\, min}$ the superconducting
state with a vortex ($m=1$) on the cylinder axis is impossible.

In Fig. 10($b$) ($m=1$), among others, the phase diagrams are depicted for
$\kappa\ge 1$. The corresponding curves (for $h_\xi>0$ and $R_\xi\gg 1$) have
the asymptote $h_\xi=1$ (or $H=H_{c2}$), common to all type-II superconductors
with $\kappa\ge 1$ (see Figs. 1 and 2). However, if $\kappa<1$, these
asymptotes depend on $\kappa$ and began deviate from the value $h_\xi=1$.
[Thus, $h_\xi\to 1$ for $\kappa=1$; $h_\xi\to 1.058$ for $\kappa=0.9$;
$h_\xi\to 1.218$ for $\kappa=0.8$; $h_\xi\to 1.416$ for $\kappa=1/\sqrt{2}$.]
This means, that in the case $m=1$ the value $\kappa=1$ is singular for
equations (1)--(6). At this value the solution passes from one branch
($\kappa\ge 1$) to another ($\kappa<1$). At the same time, the value
$\kappa=1/\sqrt{2}$ is also singular, because at $\kappa=1/\sqrt{2}$
the second branch of solutions appears (or vanishes), which describe the
edge-suppressed states [compare the behavior of the curves $h_s$ in
Figs. 2($a-e$)].

Now, we present the examples of concrete solutions of Eqs. (1)--(6),
as functions of the space coordinate $\rho=r/\lambda$ ($0\le r \le R$) for
$m=1$. (The solutions profiles for the vortex-free state ($m=0$) are shown
in [25], Fig. 2.)

In Fig. 11 the solutions are shown in the case $R_\lambda=5$, $m=1$, for
several values of $h_\lambda$. Note the curve {\it 3} in Fig. 11($a$),
which corresponds to $h_\lambda=0$ and describes a screened Meissner-type
vortex state $v_1$ ($m=1$). (In Fig. 6($a$) to this state corresponds the
point {\it 4}.)\, The curve {\it 4} in Fig. 11($a$) corresponds to the point
{\it 5} in Fig. 6($a$), where $h_\lambda=1.8254$. Here the solution branch
$v_1$ terminates, the temporal instability develops, and at $h_\lambda=1.8255$
the solution passes to new stable (static) branch $e_1$ (the edge-suppressed
state $m=1$, the point {\it 6} in Fig. 6($a$)). The curve {\it 5} in
Fig. 11($a$) corresponds to the value $h_\lambda=1.8255$ and describes the
edge-suppressed state $e_1$ with a vortex ($m=1$) at the cylinder axis (here
$\psi(0)=0$ and the order parameter is additionally strongly suppressed near
the cylinder surface).

It is expedient to trace in more details the system behavior in the states with
different $m$, when the cylinder radius $R$ is fixed, and only the external
field $H$ varies. If the superconducting region in Fig. 1($a$) is crossed along
the line $R_\lambda=2$, the states with $m=0,1,2,\cdots,11$ are possible.
These states are represented in Fig. 12 by the following dependencies:
($a$) -- the normalized free-energy difference ($\Delta g$); ($b$) -- the
magnetization ($-4\pi M_\lambda$); (c) -- the maximal value of the order
parameter ($\psi_{max}$) in the state $m$.

In  Fig. 12($a$) the free-energy curves with the adjacent $m$-values intersect
at the points, marked by the open circles. At these points the equilibrium
transitions from the state $m$ to the lower energy state $m+1$ (or reversed)
may occure. Such equilibrium transitions are acompanied by the reversible
jumps in the magnetization (see the broken vertical lines in Fig. 12($b$)).
Thus, the equilibrium magnetization curve should be reversible and have a
characteristic saw-like shape. Note also, that in the case of equilibrium
transitions the magnetization should be positive for all external fields
($-4\pi M_\lambda>0$), this corresponds to the diamagnetic susceptibility.
[The Meissner state with $m=0$ is always diamagnetic.]

However, if the metastable states with $m>0$ are also admissible (i.e. those
parts of curves in Fig. 12($a$), which lay to the left of the intersection
points), then the magnetization may turn negative within some field interval
($-4\pi M_\lambda<0$ for $m\ge 1$), this corresponds to the paramagnetic
susceptibility. The magnetization curve in this case might be irreversible
and display the hysteresis behavior.

Note also, that some of the magnetization curves in Fig.12($b$) ($R_\lambda=2$)
have smooth "tails" (for instance, at $m=0$), what indicates to the presence of
the edge-suppression effect. For larger radii ($R_\lambda > R_\Delta\sim 4$),
instead of the inflection points, the reversible jumps in the magnetization may
appear, which are analogous to those in Fig. 6($a$). Probably, such qualitative
details may be seen in the experiments with mesoscopic samples.

As is evident from Fig. 12($c$), each of the superconducting $m$-states can
exist only within a limited interval of fields, where $\psi_{max}>0$. When the
field $h_\xi$ is increased, only the last state (with $m=11$) survives. In this
state the superconducting region with
$\psi\ne 0$ is maximally pushed toward the cylinder surface, but the interior
is almost normal (see Fig. 13 below). Evidently, the state with $m=m_{max}$ is
just the state of the surface superconductivity [26,27]. In a massive cylinder
($R_\lambda\gg 1$) the last traces of the superconductivity may persist up to
the field $h_\xi=1.69$ [or, the third critical field $H=H_{c3}=1.69H_{c2}$].
At smaller $R_\lambda$ the superconductivity may persist in the fields, which
exceed $H_{c3}$ (see Fig. 1($a$)).

A peculiar field-compression effect is also present in Fig. 12. The first
superconducting solution, which appears in a field-cooled regime, has maximal
possible $m$ (the state of the surface-enhanced superconductivity). When the
external field diminishes, this state becomes impossible, but new $m-1$-state
appears, in which the superconducting region is shifted toward the center, and
simultaneously the edge-suppressed region begins to form. The state $m=4$, for
instance, has both the region of edge-suppressed superconductivity (where the
external field is practically not screened), and the region at some distance
from the center, where the field of the giant vortex ($m=4$) is compressed and
the superconductivity is also almost crushed (because at the center the order
parameter is always small, $\psi(\rho)\sim\rho^m\ll 1$). (A flux-compression
effect was discussed also in [6]).

As was mentioned, Fig. 12($b$) indicates to the possible paramagnetic
susceptibility of the superconducting cylinder in the state with $m>0$. This
topic also deserves a special comment.

The paramagnetic Meissner effect (or, Wohlleben effect) was initially observed
in [40] and extensively discussed in the literature (see, for example, Refs.
[6,8,11,15,41-43] and references therein). A number of controversial
explanations of this effect were proposed, but there is as yet no clear
understanding of its nature. This effect is observed, in particular, in a
finite-dimension samples, in the field-cooled regime, it shows the signs of
metastability, of the reentrant behavior and hysteresis [43]. It is interesting,
that all these features follow naturally from Eqs. (1)--(6) of the
Ginzburg--Landau theory.

Consider, for instance, the state $m=4$ (which is distinguished, for clarity,
by thick curves in Fig. 12). This state can not exist in small external
fields $h_\xi$, because the vortex magnetic field ($m=4, R_\lambda=2$) is
too strong for a mesoscopic sample, and can not be confined inside. However,
the imposition of a finite external field prevents the field dissipation, the
situation stabilizes  and the state $m=4$ becomes possible again in a larger
field (the reentrance effect). Evidently, it is more easy to occupy the state
$m=4$, going down from the stronger fields $h_\xi$ (in a field-cooled regime).
The negative part of the magnetization curve $m=4$ (the paramagnetic branch)
corresponds to a metastable state of higher free-energy (see Fig. 12. ($a$)),
than the state $m=3$. [Here the state $m=4$ is defined as metastable, on the
ground, that its free energy is less than the free energy of the state $m=3$.
Such metastability means, that the irreversible transition to the lower laying
state $m=3$ is possible. However, the state $m=4$ continues to be stable
relative small perturbations of its form.]

Note, that in the positive part of the magnetization curve the superconducting
currents flow, mainly, to screen out the external field. [This leads to the
diamagnetic response in Fig. 12($b$).]\, In the negative part of the
magnetization curve the role of the vortex field is more important, and the
superconducting screening currents flow, mainly, in the opposite direction.
[This means the sign reversion of the magnetic moment, because it is
proportional to the current, ${\bf M}\sim\int [{\bf j}_s{\bf r}]dv$. The
presence of positive and negative contributions to the total magnetic moment
is evident also from Fig. 11($f$), where the current density may be positive or
negative). At the point $M_\lambda=0$ the currents, which screen the internal
and external fields, counterbalance each other. Thus, according to the
Ginzburg--Landau theory, the paramagnetism present in Fig. 12, has purely
electrodynamic explanation.

Now we shall demonstrate, how the order parameter vanishes in approaching the
critical curves $h_c^{(m)}>0$ in Figs. 1 and 2 [the curves $h_c$ are supplied
with an additional index $(m)$]. The order parameter is small in the vicinity
of $h_c^{(m)}$. In Fig. 13 the degenerated solutions (i.e. those with
$\psi_{max}\ll 1$) are shown for $R_\lambda=2$, $\kappa=2$ and different $m$.
[The curves in Fig. 13 are normalized to unity, with true values
$\psi_{max}\ll 1$.]\, In the origin of coordinates the order parameter behaves
as $\psi(\rho)\sim\rho^m$, in agreement with the theory (see, for instance,
[29,35]). At $\rho\gg 1$ the order parameter is also small (here the
edge-suppressed state $e_m$ is formed). For all degenerated solutions,
depicted in Fig. 13 (with $\psi_{max}\ll 1$), the field $B(r)\approx H$.
In this case Eqs. (1),(2) can be linearized and reduced to one linear equation
for the function $\psi$ [26,27]. The solutions of this equation may be found
analitically, in terms of the Kummer functions (see, for instance, [6,8]).
[Notice, however, that the normalization factor in the linear equation for
$\psi$ remains arbitrary, and to find the true value $\psi_{max}$ it is
necessary to solve the full system of nonlinear equations (1)--(6).]\,
The critical fields $h_c^{(m)}>0$ in Fig. 1, which we found by solving the
full system of equations (1)--(6), may be obtained more easily on the base of
a linear equation (when $\psi_{max}\ll 1$), as was done in [6,8]. [However,
when approaching the phase boundaries $h_c^{(m)}$ for negative $h_\xi<0$, or
to find the critical fields $h_s^{(m)}$, when the edge-suppressed state forms
(see Figs. 1 and 2), but the order parameter $\psi_{max}$ remains finite, it
is necessary to use the full system of equations (1)--(6).]

It is evident from Fig. 13, that even the degenerated solutions (with
$\psi_{max}\ll 1$) display the presence of the edge-suppresssion effect. If
$m=m_{max}$, the order parameter is maximal at the cylinder surface. For
$m<m_{max}$ the order parameter begins being suppressed at the boundary. When
the field $H$ diminishes, only the states with relatively small $m$ survive
[with the inflexion points on the magnitization curves, see Figs. 12($b,c$)],
and finally the edge-suppressed solutions would form [analogous to those, shown
in Fig. 11, with the characteristic jumps, as in Figs. 3 and 6].

Finally, it is of interest to note, that the transitions between states with
different $m$ are also possible in type-I superconductors (see also [35,11]).
This is demonstrated by Fig. 14. In Figs. 14($a,b$) the profiles of the
self-consistent solutions $\psi(r)$ and $b(r)$ are shown for type-I
superconductor ($\kappa=0.5$, $R_\lambda=4$) in the fields: $h_\xi=1.4892$
($m=0$); $h_\xi=2.0344$ ($m=1$); $h_\xi=2.1084$ ($m=2$). These $h_\xi$-values
are in the immediate vicinity of the corresponding critical curves $h_c^{(m)}$
in Fig. 1($b$): if $h_\xi$ is increased by $1\cdot 10^{-4}$ [i.e., if the
critical curves in Fig. 1($b$) are crossed along the line $R_\lambda=4$], the
transitions to the normal state occures ($\psi(r)=0$, $B(r)=H$).

As is clear from Fig. 14($a$), in type-I superconductors the transition to the
normal state is of the first order (by a jump from a finite value
$\psi_{max}^*$). However, the jump amplitude (i.e. the value $\psi_{max}^*$ at
the transition point) depends on parameters $m$, $\kappa$ and $R_\lambda$ (or
$R_\xi=R_\lambda/\kappa$). The jump amplitude, $\psi_{max}^*(R_\lambda)$, is
depicted in Fig. 14($c$) for $\kappa=0.1$ and $\kappa=1/\sqrt{2}$ in the case
$m=0$. As follows from Fig. 14($c$), the value $\psi_{max}^*$ diminishes, when
$R_\lambda$ diminishes, with the jump amplitude vanishing at the point
$R_\Delta$. Thus, the points $R_\Delta$ on the critical curves in Figs. 1($b$),
10($a$) and 14($c$) divide two regions of the phase transitions in the case of
type-I superconductors: for larger $R_\lambda>R_\Delta$ the first order phase
transition to the normal state occures; for smaller $R_\lambda<R_\Delta$ one
has the second order phase transition. At $R_\lambda=R_\Delta$ both transitions
became indistinguishable, because the jump amplitude $\psi_{max}^*$ vanish.
[Remind, that in type-II superconductors, with $\kappa>1/\sqrt{2}$, the
phase transition to the normal state is always second order, regardless of
the specimen size.]\, The critical size $R_\Delta$, which separates the regions
of the first and second order phase transitions in the case of type-I
superconductors, may also be found from different considerations [1,32,36,38].
Fig. 14($c$) shows, that this size depends slightly on $\kappa$.

\section{Comments and conclusions}

In conclusion, the main results of the present investigation are formulated.

The self-consistent solutions of a nonlinear system of Ginzburg--Landau
equations, which correspond to the fixed values of the magnetic quantum number
$m$ and have a cylinder symmetry (i.e. depend only on the radial coordinat $r$,
but not on the azymuthal angle $\varphi$), are studied.

It is found, that for type-II superconducting cylinder of finite radius
($R>R_\Delta$), placed in sufficiently weak magnetic field $H$, the order
parameter $\psi(r)\approx\rm{const}$ near the cylinder surface, and behaves as
$\psi(r)\sim r^m$ near the center (the giant-vortex state). If the field is
increased ($H>H_s^{(m)}$), the giant-vortex state changes abruptly and acquires
new form. (A first order phase transition to the edge-suppressed state occures.)
In this new state the order parameter is strongly suppressed near the cylinder
edge, with the magnetic field penetrating practically without screening into
some layer in the vicinity of the surface. At the center the behavior
$\psi(r)\sim r^m$ remains. This transition to the edge-suppressed form is
acompanied by jumps in the magnetization and in the total magnetic flux,
confined in the cylinder. The properties of such edge-suppressed solutions
(which depend on the parameters $m,H,\kappa,T$) are studied. It is shown, that
the edge-suppressed states exist only in type-II superconductors, but the
giant-vortex states are possible both in type-I and type-II superconductors.
The phase boundaries are found, inside which the giant $m$-vortex solutions can
exist. In particular, the characteristic radius of the cylinder is found,
$R_\Delta$, which separates the regions of the first and second order phase
transitions between the superconducting and normal states.

The behavior of the magnetization ($M$) and of the total magnetic flux
($\Phi_1$) as functions of the external field ($H$) is found, with account of
the jumps, which acompany the formation of the edge-suppressed states. It is
demonstrated, that some maximal value $m=m_{max}$ exist, for which the
edge-suppressed state degenerates into the usual state of the surface
superconductivity (with nucleation field $H\approx H_{c3}$ [26,27]).
The Gibbs free energies of different axially symmetrical superconducting
states are compared, and the problem of paramagnetic effect, observed in
mesoscopic samples, is shortly discussed. It is found, that the paramagnetic
effect is possible in metastable vortex states and it may be attributed
to the disbalance of the superconducting currents, which screen the external
field $H$ and the field of the vortex from entering the sample interior.

As was pointed out earlier [25], according to the Ginzburg--Landau theory, two
competing mechanisms of the external field penetration into the superconducting
cylinder exist -- in a form of vortices [2--15,30], and in a form of the
edge-suppressed layer. The question, which of these mechanisms is more
favorable energetically, may be answered only after the full comparison is made
of the Gibbs free energies for the arbitrary multy-vortex configurations in a
finite radius cylinder. This difficult theoretical problem (as well as the
generalisation to other geometries) is left outside the scope of the present
investigation. Some of the edge-suppressed states, described above,
are metastable, but they may be observable in a specially arranged experiments.
We beleve it is expedient to draw attention to this possibility.

\section{Acknowledgments}

We are gratefull to V. L. Ginzburg for the interest to this work and valuable
discussions.

\centerline{\bf Figures captions}

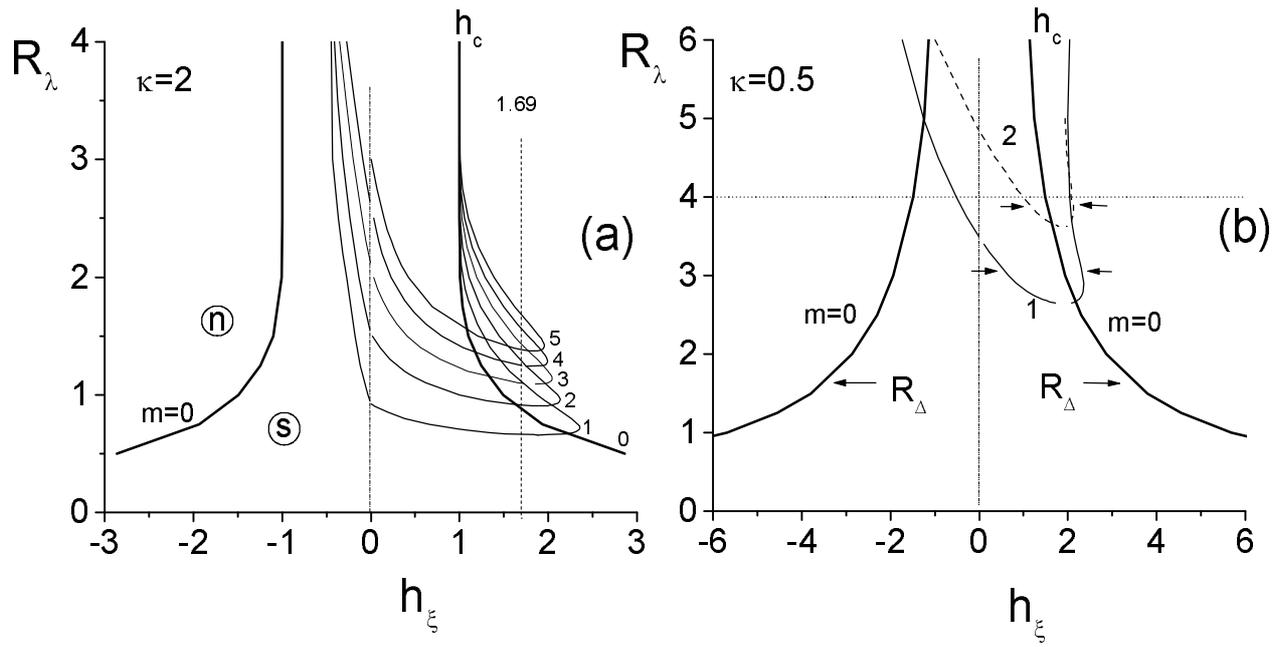
Fig. 1.$\,\, (a)$ -- The phase diagram of type-II superconductor ($\kappa=2$).
The numerals at the curves -- the values of the magnetic quantum number $m$.
The value $h_\xi=1$ corresponds to the critical field
$H_{c2}=\phi_0/(2\pi\xi^2)$. The value $h_\xi=1.69$ corresponds to the maximal
field $H_{c3}=1.69H_{c2}$, in which the surface superconductivity is still
possible in macroscopic specimens.\  ($b$) -- The phase diagram of type-I
superconductor ($\kappa=0.5$, $m=0,\, 1,\, 2$). The arrows and the letter
$R_\Delta$ at the curves denote the points, where the first and second order
phase transitions to the normal state became idistinguishable (see the text).\\

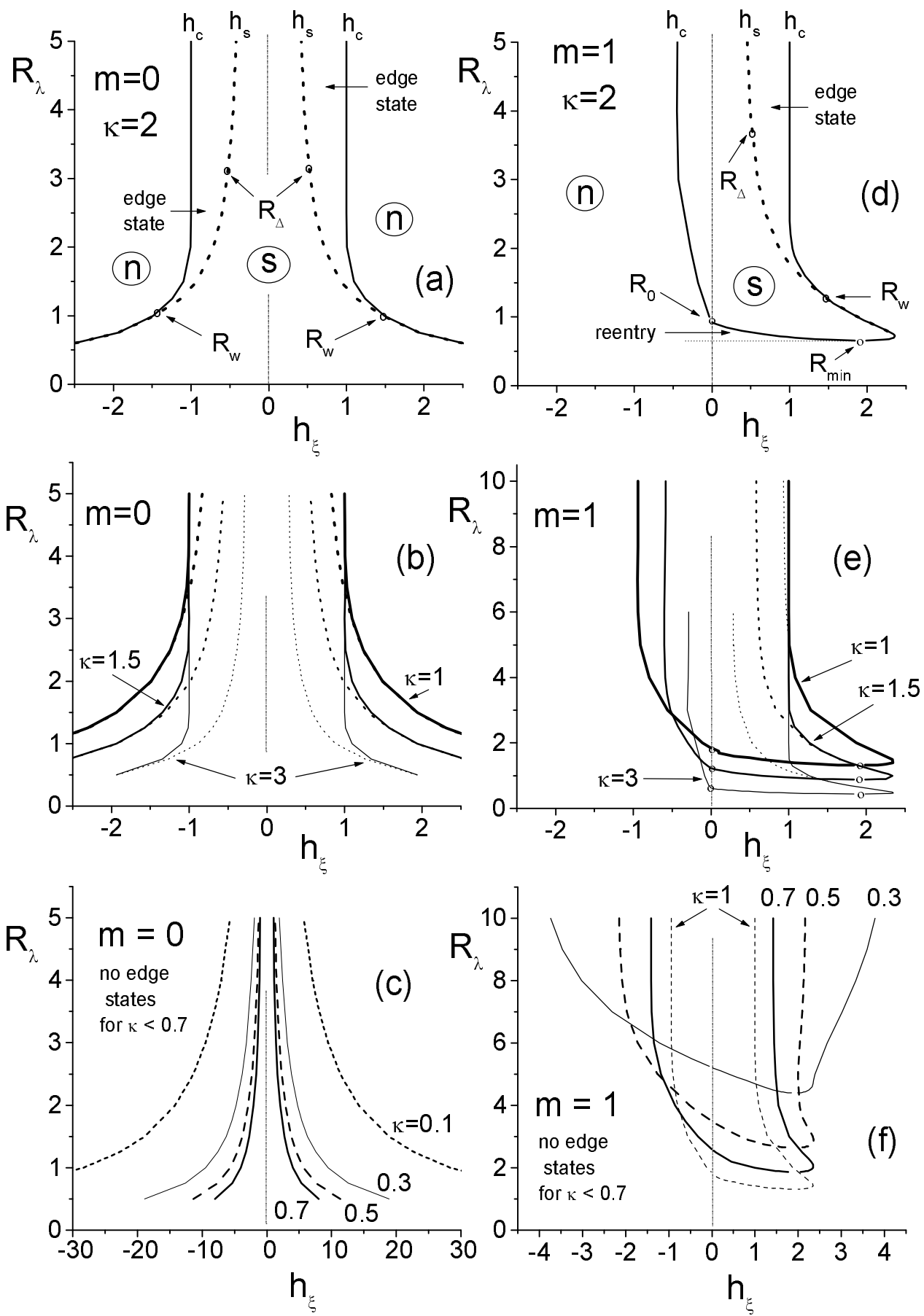
Fig. 2. ($a$) -- The case $m=0$, $\kappa=2$. The phase curve $h_c$ divides
superconducting ($s$) and normal ($n$) states of the cylinder.
When the representation point $(R_\lambda,h_\xi)$ crosses the line $h_s$ (the
dotted line), a new edge-suppressed solution forms, with suppressed order
parameter $\psi_1$ (see Fig. 3($f$)). No edge-suppressed states exist for
$R_\lambda<R_w$.
($d$) --  Analogously, for the case $m=1$, $\kappa=2$. When $R_\lambda<R_0$,
the field $h_\xi>0$ stimulates the superconducting state, so the reentry
superconductivity is possible.
($b$) and ($e$) -- Analogously, for different $\kappa\ge 1$.
($c$) and ($f$) -- The phase curves of type-I superconductor (for various
$\kappa\le 1/\sqrt{2}$). The edge-suppressed states in type-I superconductors
are abscent.\\

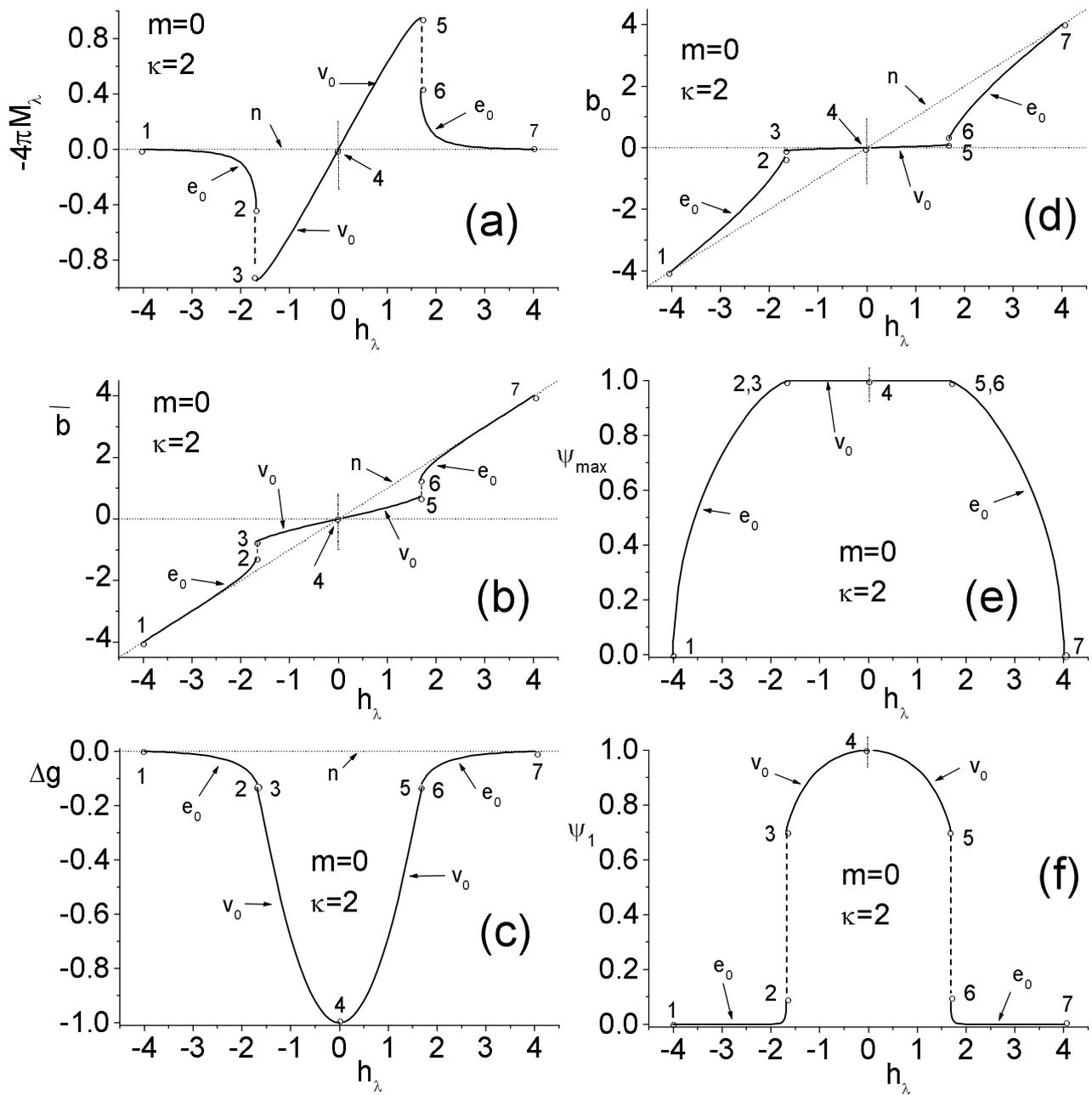
Fig. 3. The case $m=0$, $\kappa=2$, $R_\lambda=5$. ($a$) -- The cylinder
magnetic moment, $M_\lambda=M/H_\lambda$.
($b$) --The average magnetic field in the cylinder, $\overline{b}$ (or, the
total magnetic flux in the system, $\phi_1={1\over 2}\rho_1^2\overline{b}$ (8)).
($c$) -- The normalized difference of free energies, $\Delta g$, (10).
($d$) -- The magnetic field at the cylinder axis, $b_0$.
($e$) -- The maximal value of the order parameter, $\psi_{max}$.
($f$) -- The order parameter at the cylinder surface, $\psi_1$. The letter
$v_0$ denotes the Meissner state (vortex-free, $m=0$). The letter $e_0$ denotes
the edge-suppressed state (vortex-free, $m=0$). The transition from one branch
of the solution to another takes place between points {\it 2,3}, or {\it 5,6}.
(The solution profiles for the case $R_\lambda=5$, $m=0$, $\kappa=2$ are
depicted in [17], Fig.2.)\\

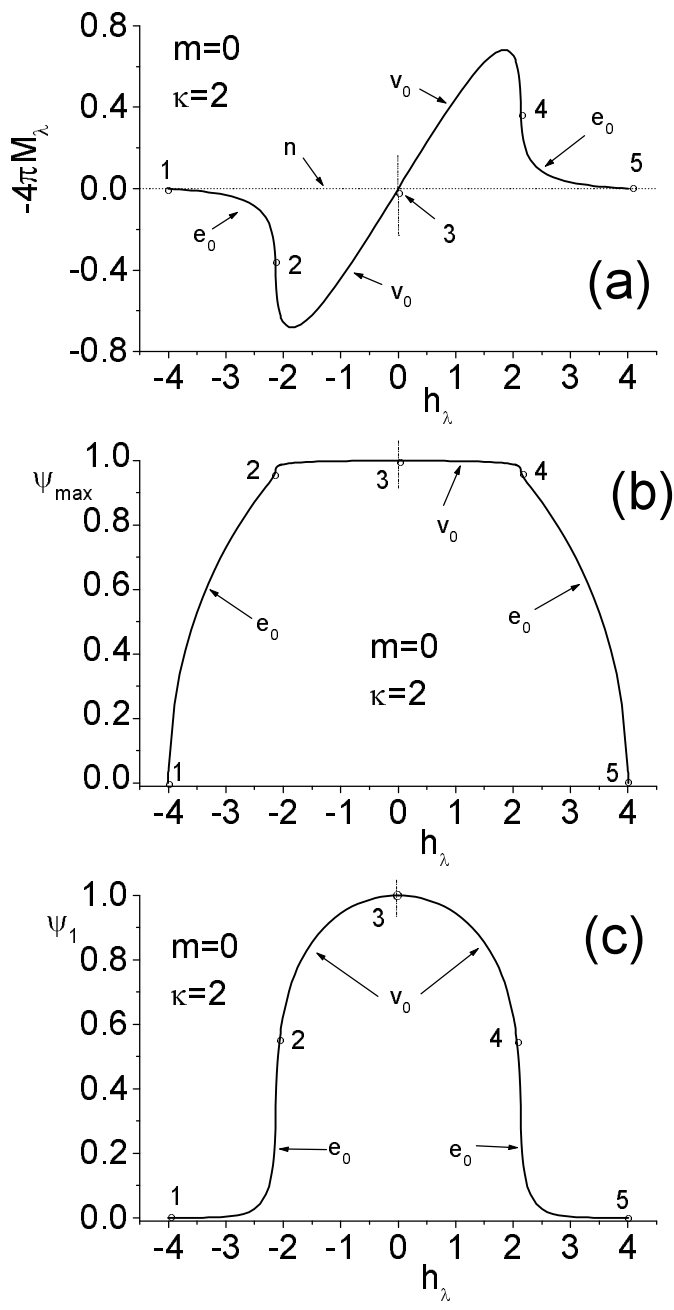
Fig. 4. Analogous to Fig. 3, but for the case $m=0$, $\kappa=2$, $R_\lambda=3$.\\

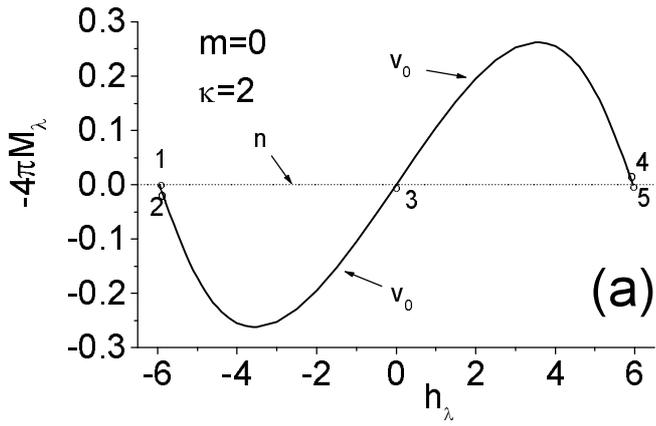
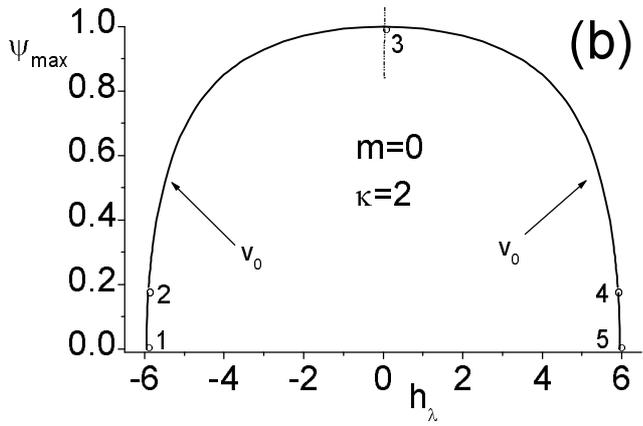
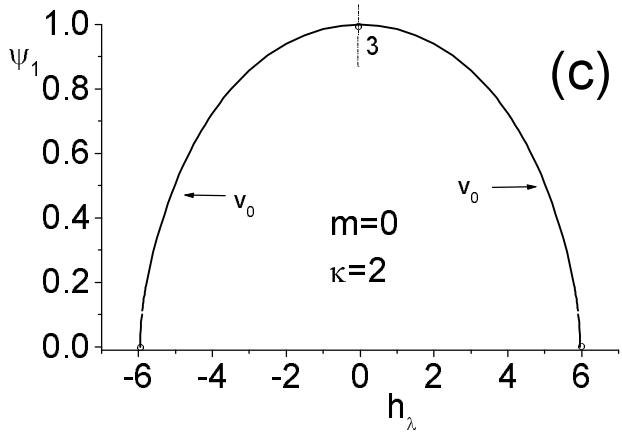
Fig. 5. Analogous to Fig. 3, but for the case $m=0$, $\kappa=2$, $R_\lambda=1$.\\

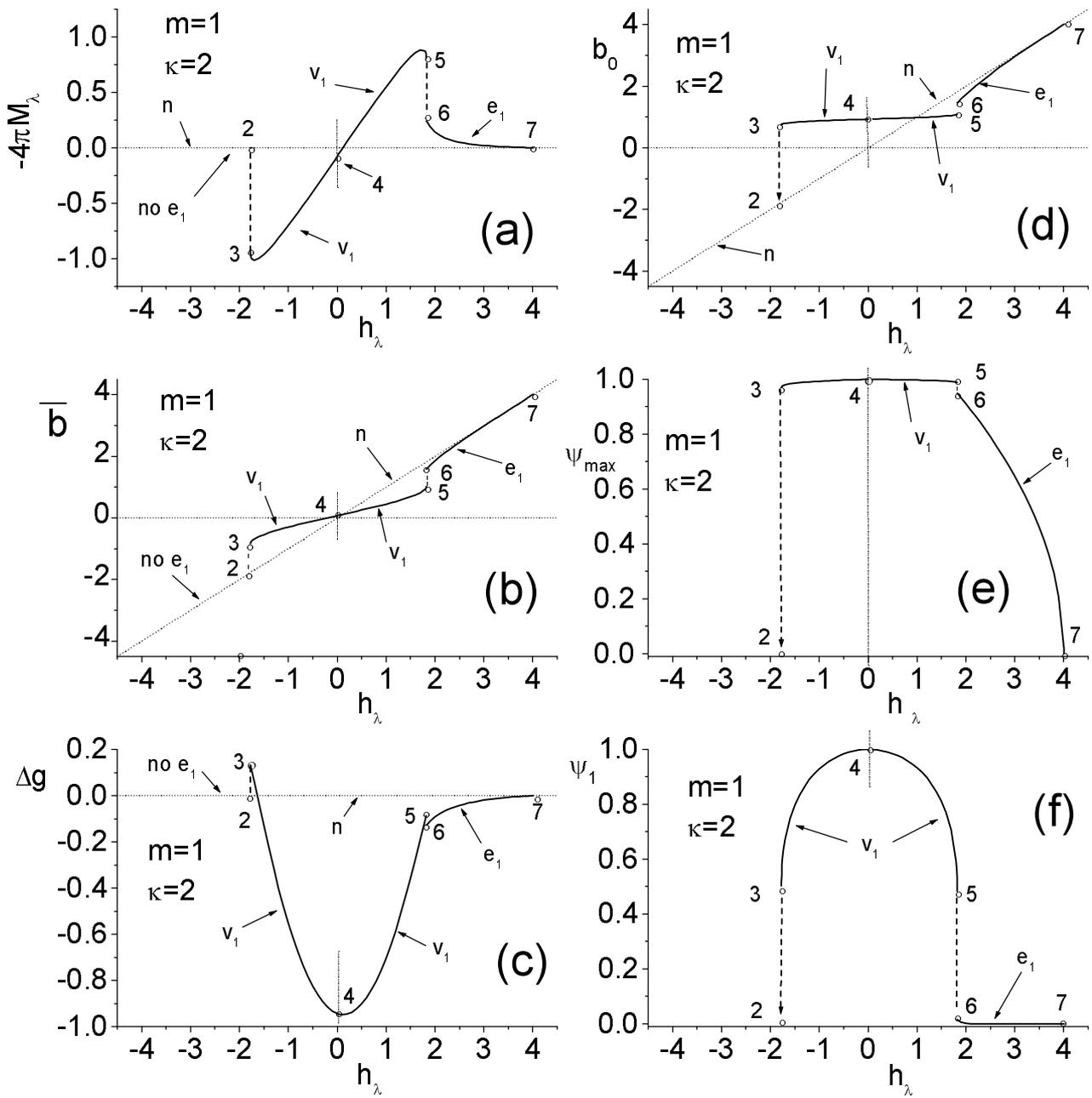
Fig. 6. The same, as in Fig. 3, but for $m=1$, $\kappa=2$, $R_\lambda=5$.
The letter $v_1$ denotes the Meissner-type vortex state (with a vortex $m=1$
at the cylinder axis). The letter $e_1$ denotes the edge-suppressed state
with the vortex ($m=1$) at the axis. (The solutions profiles in the case $m=1$,
$\kappa=2$, $R_\lambda=5$, are shown below in Fig. 11.)\\

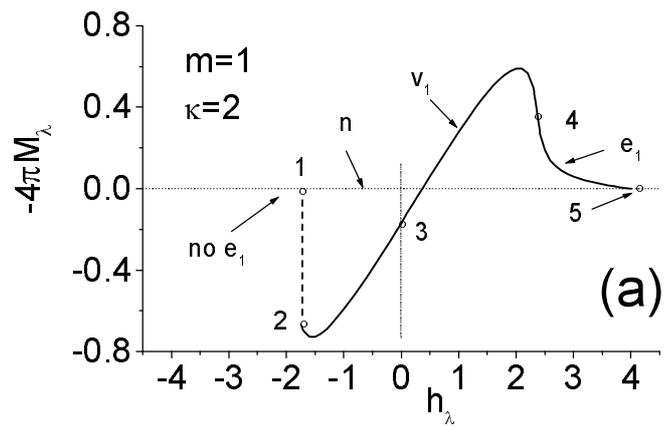
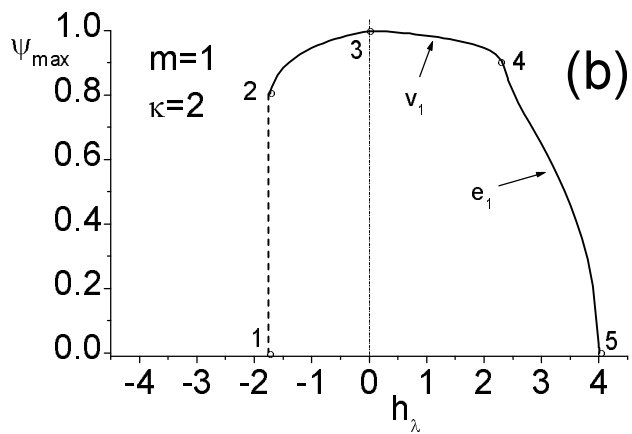
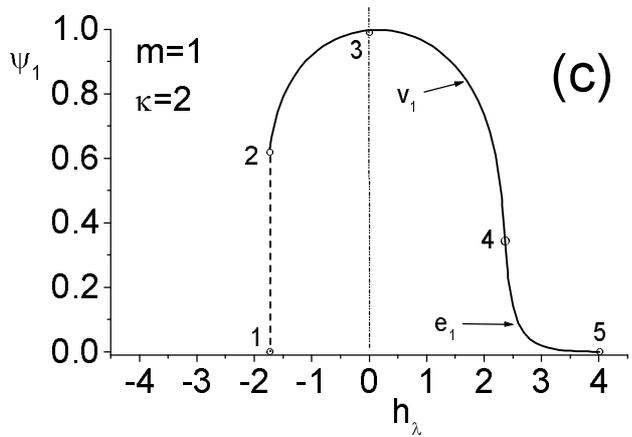
Fig. 7. Analogous to Fig. 3, but for the case $m=1$, $\kappa=2$, $R_\lambda=3$.\\

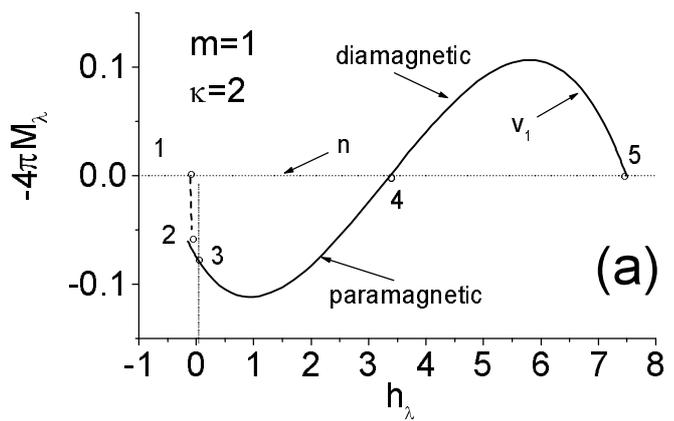
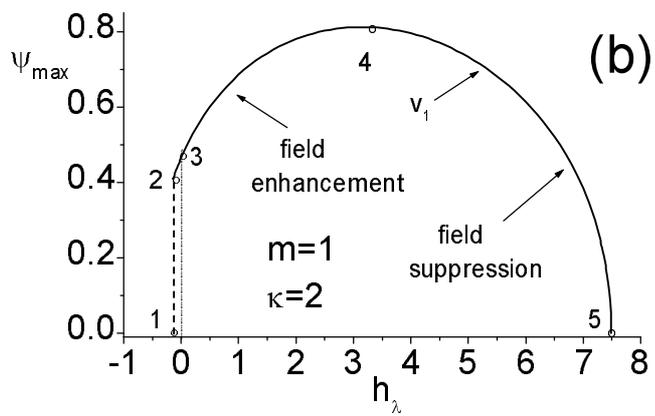
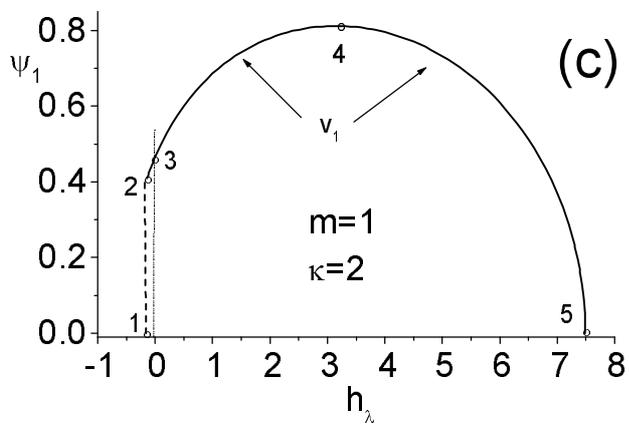
Fig. 8. Analogous to Fig. 3, but for the case $m=0$, $\kappa=2$, $R_\lambda=1$.\\

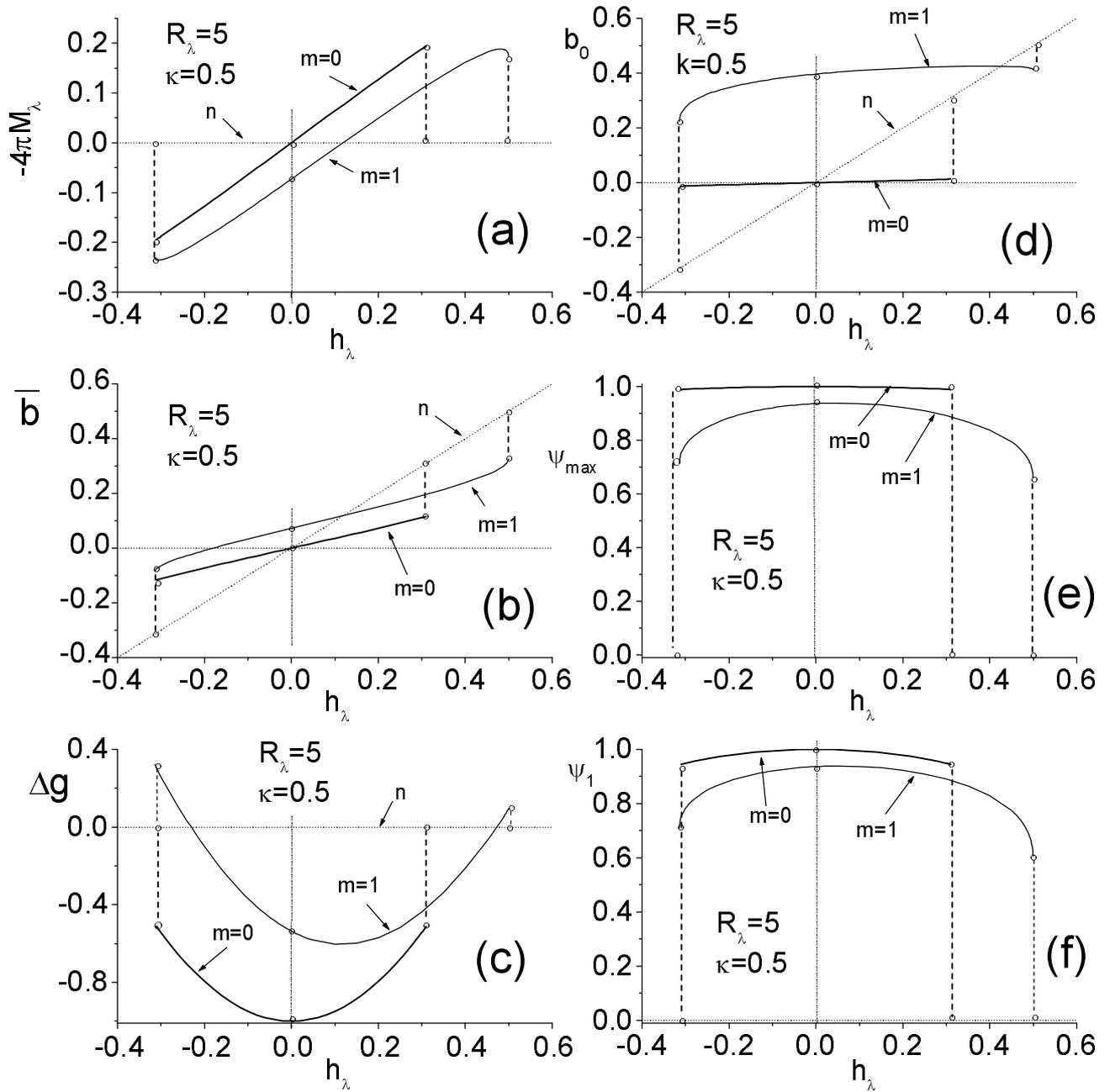
Fig. 9. The same, as in Fig. 3, but for type-I superconductor
($\kappa=0.5$, $R_\lambda=5$, $m=0$ and $m=1$). There are no edge-suppressed
states and no "tails" in the magnetization for type-I superconductors. \\

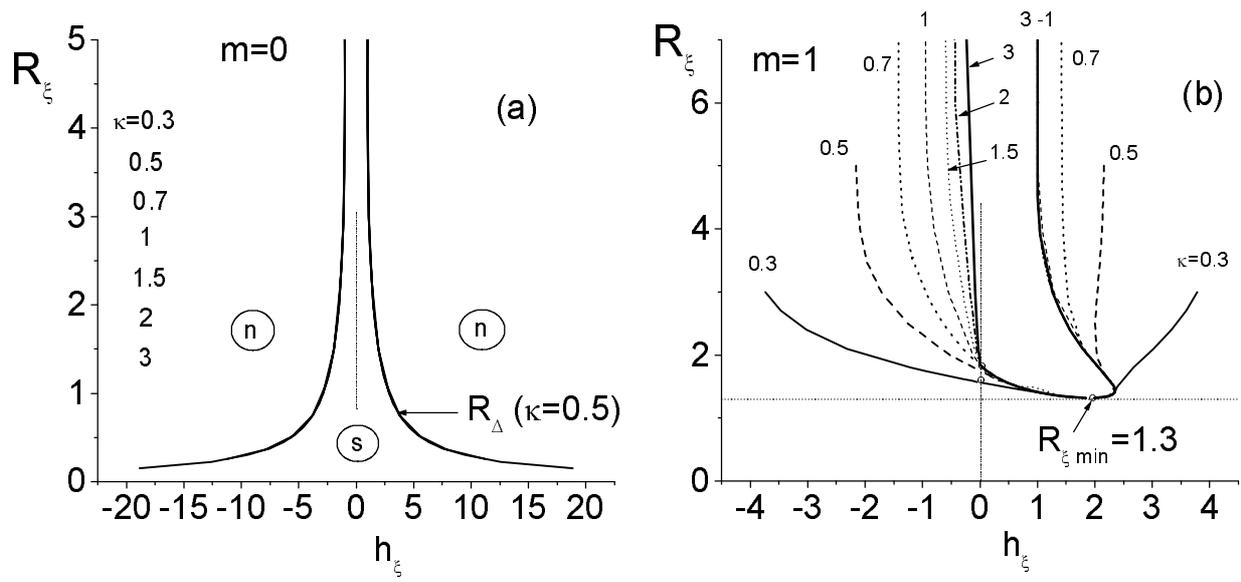
Fig. 10. ($a$) -- The phase diagram for $m=0$ and different $\kappa$. The
arrow and letter $R_\Delta$ devide the regions of the first and second order
phase transitions to the normal state in the case of type-I superconductors.
For $m=0$ the curves with different $\kappa$ [see numbers in Fig. 10($a$)]
practically coincide.
($b$) -- The phase diagram for $m=1$ and different $\kappa$. For
$R_\xi<R_{\xi\,min}$ the superconducting state with the vortex ($m=1$) at the
axis is impossible.\\

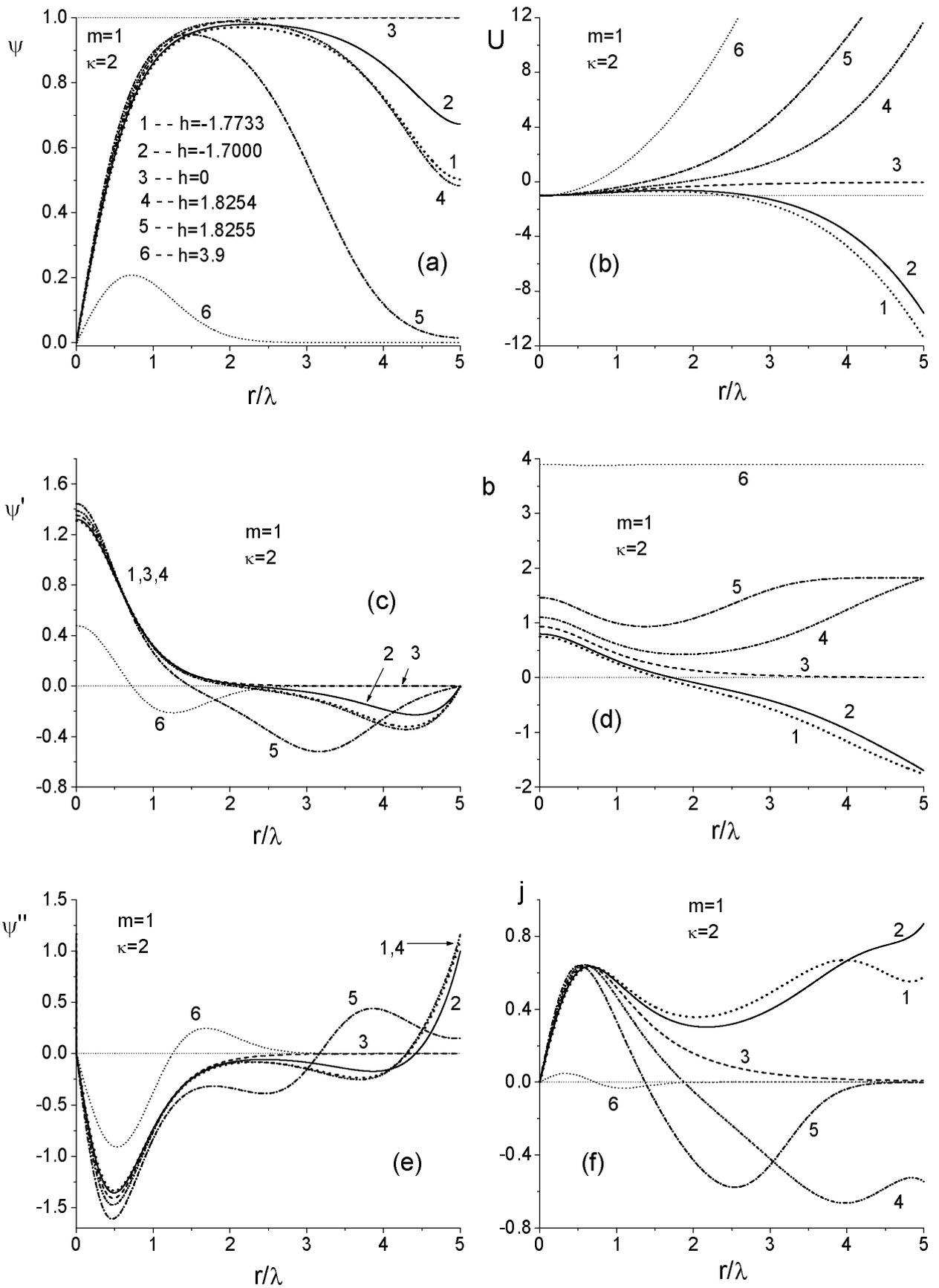
Fig. 11. The coordinate dependence of solutions for $m=1$, $R_\lambda=5$,
$\kappa=2$ and different fields (see the values in Fig. 11($a$).
($a$) -- The order parameter, $\psi(\rho)$; $\rho=r/\lambda$.
($b$) -- The derivative, $d\psi/d\rho$.
($c$) -- The second derivative, $d^2\psi/d\rho^2$.
($d$) -- The field potential, $U(\rho)$.
($e$) -- The normalized magnetic field, $b(\rho)$.
($f$) -- The normalized current, $j(\rho)$.   \\

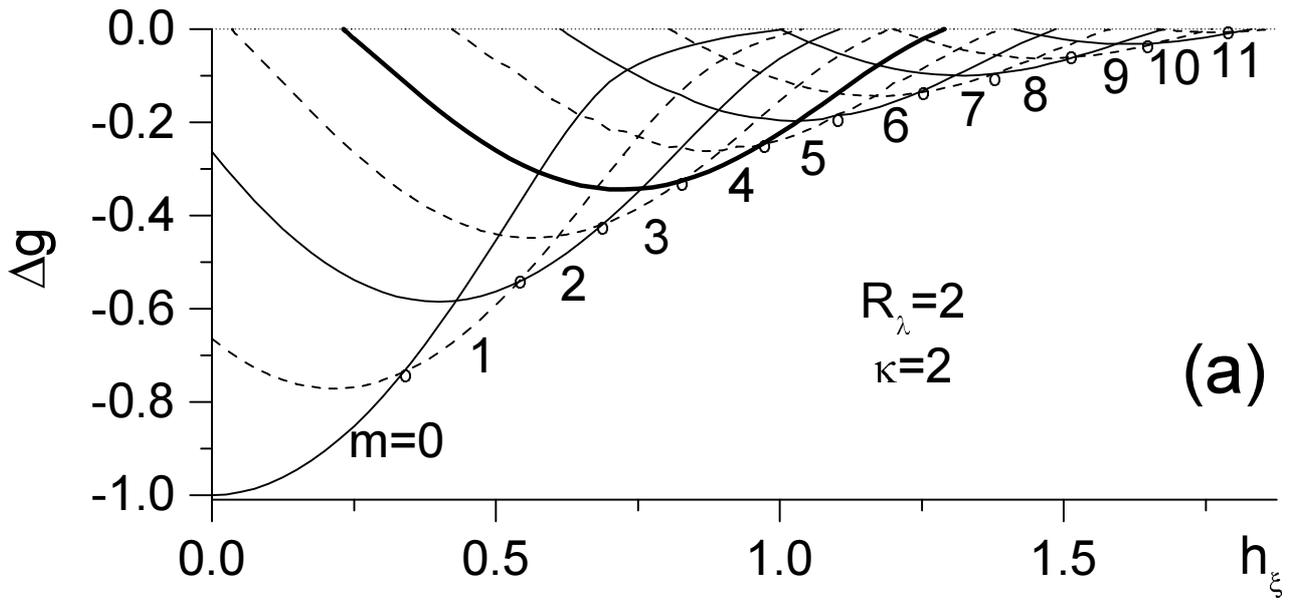
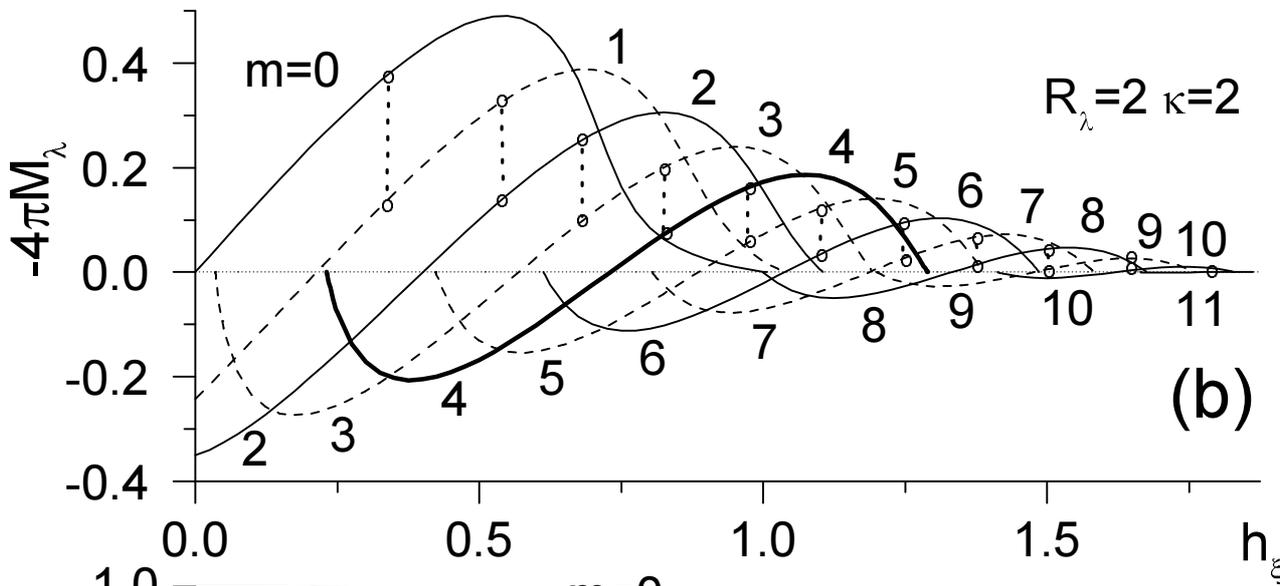
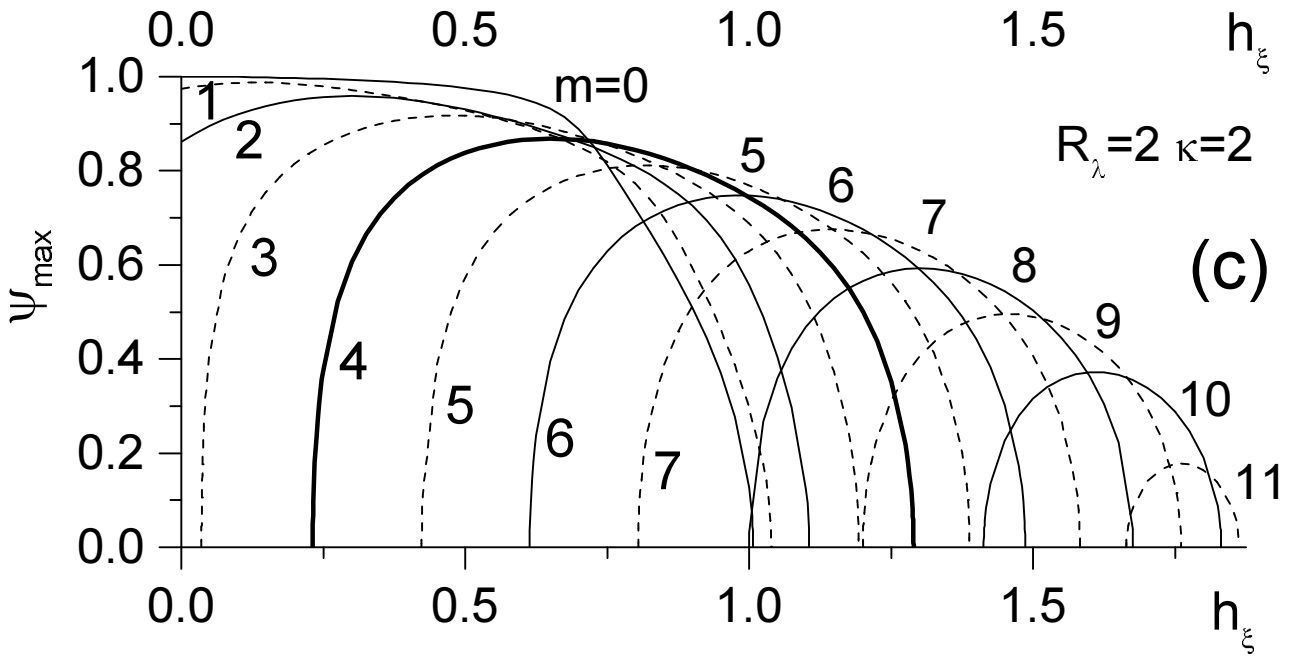
Fig. 12. ($a$) -- The free energy, $\Delta g$; ($b$) -- the magnetic moment,
$(-4\pi M_\lambda)$ and ($c$) -- the maximal value of the order parameter,
$\psi_1$, for the cylinder with $R_\lambda=2$, $\kappa=2$ and various $m$, as
functions of $h_\xi$.    \\

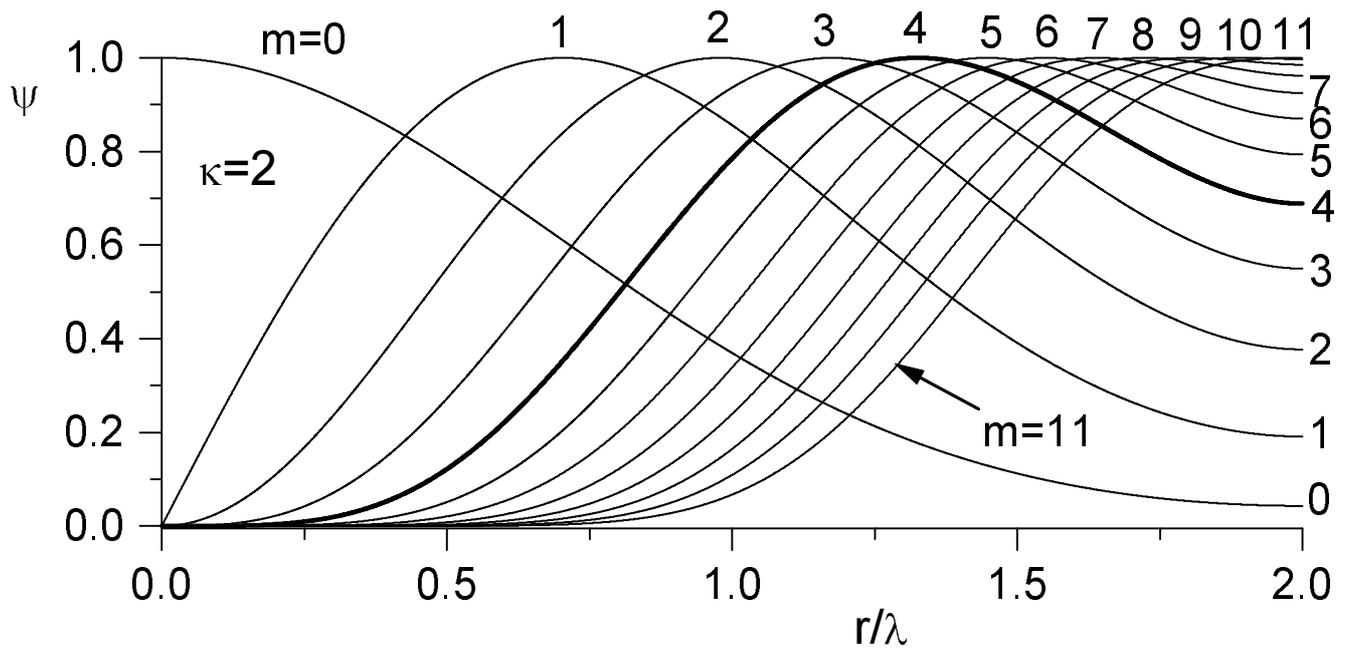
Fig. 13.  The order parameter profiles in a case of degenerated states
($\psi_{max}\ll 1$) for $R_\lambda=2$, $\kappa=2$ and various $m$. (All the
curves represent the self-consistent solutions of Eqs. (1)--(6), but they
are normalized to unity.) With $m$ increasing, the value $\psi_{max}$
approaches the boundary $r=R$. The last of the degenerated states ($m=11$),
which is maximally pushed toward the boundary, corresponds to the state of the
surface superconductivity [26,27].\\

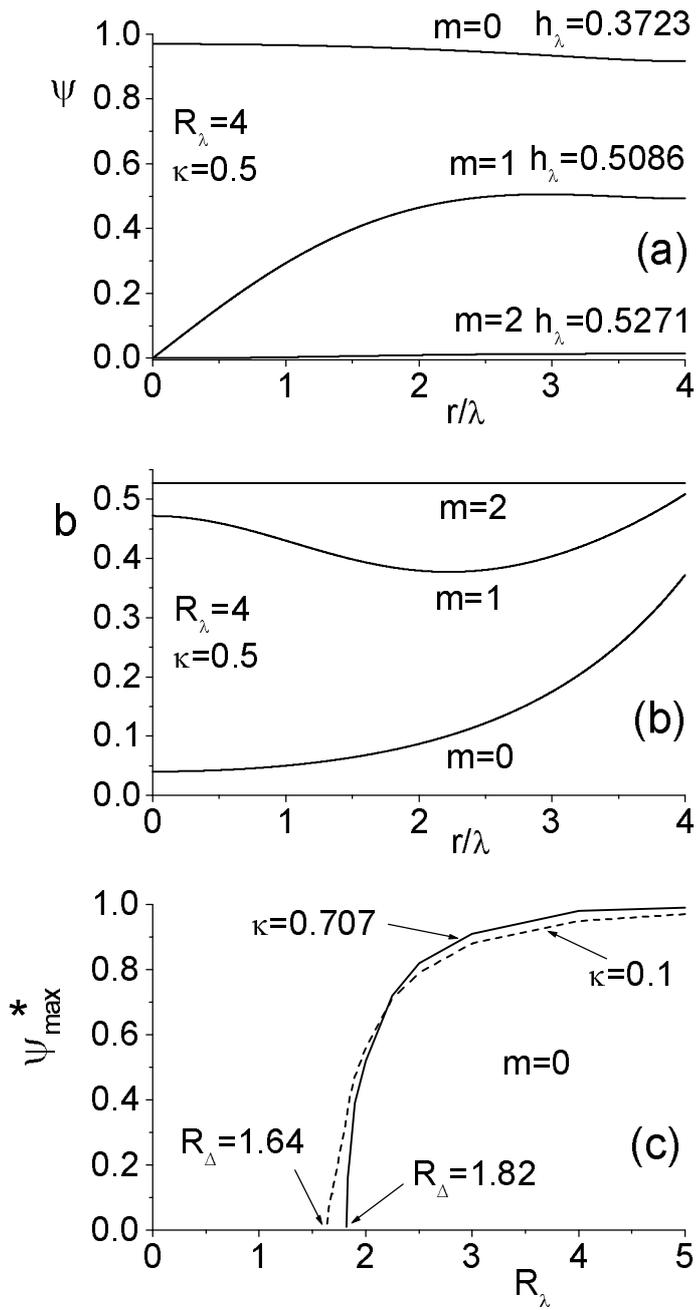
Fig. 14. ($a$) -- The order parameter profiles in a case of type-I
superconductor ($\kappa=0.5$, $R/\lambda=4$). The external fields (see the
numerals at the curves) are in the immediate vicinity of the critical boundaries
$h_c^{(m)}$ in Fig. 1($b$). If the field $h_\xi$ is increased by
$1\cdot 10^{-4}$, the jump transition to the normal state occures.
($b$) -- The corresponding magnetic field profiles.
($c$) -- The jump amplitude, $\psi_{max}^*$, as function of $R_\lambda$ for
$\kappa=0.1$ and $\kappa=1/\sqrt{2}$. [The jump amplitude $\psi_{max}^*$ is
the value of $\psi_{max}$ in a field, which just preceeds the first order phase
transition to the normal state. The value $R_\Delta$, where the jump amplitude
vanish (see Figs. 1($b$), 10($a$), 14($c$)) depends on $\kappa$.] \\

\end{document}